\documentclass[journal]{IEEEtran}
\ifCLASSINFOpdf
\else
\fi
\usepackage{subfigure}
\usepackage[dvips]{graphicx}
\usepackage{multirow}
\usepackage[square,comma,sort&compress,numbers]{natbib}
\usepackage{amssymb}
\usepackage[cmex10]{amsmath}
\DeclareMathSizes{10}{9}{8}{6}
\begin{document}
\title{Distributed Raptor Coding for Erasure Channels: Partially and Fully Coded Cooperation}
\author{Mahyar~Shirvanimoghaddam,~\IEEEmembership{Student~Member,~IEEE,}
               Yonghui~Li, ~\IEEEmembership{Senior~Member,~IEEE,}
               Shuang~Tian,~\IEEEmembership{Member,~IEEE,}
                      and Branka~Vucetic,~\IEEEmembership{Fellow,~IEEE}
\thanks{The paper has been accepted for publication in IEEE Transaction on Communications. Manuscript received September 25, 2012; revised March 7, 2013 and June 3, 2013. The editor coordinating the review of this paper and approving it for publication was M. Xiao.

The material in this paper was presented in part at the 2012 IEEE International Symposium on Information Theory, Cambridge, MA, July 2012, and in part at the 2012 IEEE Global Communications Conference, Anaheim, CA, December 2012.

The authors are with the Center of Excellence in Telecommunications,
School of Electrical and Information Engineering, University of Sydney,
Sydney, NSW 2006, Australia (e-mail: \{mahyar.shirvanimoghaddam, yonghui.li, shuang.tian, branka.vucetic\}@sydney.edu.au).

This work was supported by the Australian Research Council (ARC) under Grants DP120100190, FT120100487, LP0991663, DP0877090.}}
\maketitle
%\doublespacing
\begin{abstract}
In this paper, we propose a new rateless coded cooperation scheme for a general multi-user cooperative wireless system. We develop cooperation methods based on Raptor codes with the assumption that the channels face erasure with specific erasure probabilities and transmitters have no channel state information. A fully coded cooperation (FCC) and a partially coded cooperation (PCC) strategy are developed to maximize the average system throughput. Both PCC and FCC schemes have been analyzed through AND-OR tree analysis and a linear programming optimization problem is then formulated to find the optimum degree distribution for each scheme. Simulation results show that optimized degree distributions can bring considerable throughput gains compared to existing degree distributions which are designed for point-to-point binary erasure channels. It is also shown that the PCC scheme outperforms the FCC scheme in terms of average system throughput.
\end{abstract}
\IEEEpeerreviewmaketitle

\section{Introduction}
\IEEEPARstart{T}{he} broadcast nature of wireless transmission enables users in the network to overhear transmissions of other surrounding users. As a result, each user can help other users in forwarding their messages. A spatial diversity can be achieved by each user through such a cooperative transmission process \cite{UserCoop1,UserCoop2}. One of the most important user cooperation strategies is coded cooperation \cite{DivThroCodedCoop} which can achieve both spatial diversity and cooperative coding gains.
Various coded cooperation schemes based on convolutional, turbo, space time and LDPC codes have been proposed \cite{LiTurbo,LiSpaceTime,CodedWireless,LDPCRelay}.

Recently, coded cooperation schemes based on rateless codes have attracted considerable interests. Unlike other error correcting codes, rateless codes do not need to be designed for a predetermined rate, and thus the code design process is much easier. There are several works in the literature that have used rateless codes to achieve cooperative diversity. In \cite{RatelessRelay,FountainRelay,rateHalfDup}, Raptor codes have been applied to a simple three node relay network and the work was further extended to the general M-relay system in \cite{DynamicDF,BufferedRateless,RaptorDownlink,FountainrelayMolisch,LowNikjah}. Also, in \cite{DisUEP}, a distributed rateless code with an unequal error protection (UEP) property has been proposed. Different data importance levels necessitate the design of distributed rateless codes with different error probabilities for different sources. In \cite{DisUEP,StreamUEP,UEP,sejdinovic2009expanding}, it has been shown that rateless codes can be efficiently applied to distributed networks to achieve UEP property.

The focus of this paper is on the design of a practical high throughput user cooperation strategy for the cooperative multiple access channel (CMAC), based on Raptor codes \cite{Raptor} and the assumption that the channels face erasure with specific erasure probabilities. Such a cooperative strategy has been previously proposed in \cite{NewRatelessCodedCoop} based on existing point-to-point degree distributions, where only one user helps the other user to forward its message. Since the cooperative transmission process in coded cooperation will change the degree distribution of the overall codeword at the destination, existing point-to-point degree distributions will not be optimal any more, thus the scheme in \cite{NewRatelessCodedCoop} performs poorly in practice.

In this paper, two coded cooperation strategies are developed for a general M-user CMAC, referred to as a fully coded cooperation (FCC) scheme and a partially coded cooperation (PCC) scheme. The FCC scheme \cite{MahyarISIT} is quite similar to the conventional rateless coded cooperation \cite{CoopLowMACtrans}, where the overall transmission is divided into a broadcast and a cooperative phase. In the broadcast phase,  each user keeps on transmitting its own coded symbols until it completely recovers at least one other user's message. When a user successfully decodes at least one of the other users' messages, it independently starts the cooperation phase and generates coded symbols from its own message and other users' messages successfully decoded in the broadcast phase, and sends them to the destination. However, in the FCC scheme the cooperation cannot start until the user has successfully decoded at least one partner's message. This will lead to a low transmission efficiency  especially when the inter-user channels are poor. To overcome this problem, we proposed a PCC scheme in \cite{MahyarGlobe} to fully exploit coded cooperation. In the PCC scheme, each user tries to partially decode other users' messages from the received coded symbols in the current and previous time frames. Thus the cooperation is performed throughout the whole transmission process and this will significantly improve the overall system performance.

Since conventional degree distributions are designed only to guarantee a good full decoding performance; they will not perform well when the number of received coded symbols is smaller than that of information symbols \cite{ImprovedInterPerfRateless}, leading to a very poor performance when applied to PCC. Partial decoding performance of rateless codes in terms of recovery rate is defined as the ratio of the number of information symbols recovered in partial decoding to the total number of information symbols \cite{IntPerfRateless}, and it has been widely discussed in \cite{IntPerfRateless,RateIntermediate,Growth,ImprovedInterPerfRateless}. It has been shown in \cite{RateIntermediate} that when the partial decoding  performance is increased, the fully decoding performance will degrade. Although Growth codes \cite{Growth,ImprovedInterPerfRateless} achieves both good full and partial decoding performance, they require a large number of feedback from the destination; thus not appropriate for the PCC scheme.

In this paper, we will design the optimal degree distributions for both PCC and FCC schemes in a general M-user cooperative multiple access channel. We optimize the degree distribution in a way that when the destination knows a part of users' messages, other parts can be recovered at a minimum overhead. We will follow the similar linear optimization method originally proposed for degree optimization in \cite{CoopLowMACtrans}, where the degree distribution is determined to decode all users' messages with minimum overhead when a number of users' messages are known at the destination. The performance of PCC and FCC scheme are then analyzed by using AND-OR tree analysis and verified by simulation results.  Both analytical and simulation results show that the PCC scheme outperforms the FCC scheme in terms of the average system throughput. Moreover, the optimized degree distributions are compared with existing degree distributions originally designed for point-to-point transmission \cite{Raptor}. It is shown that the optimized degree distributions considerably outperform existing degree distributions for both FCC and PCC schemes in terms of the average system throughput.

Throughout the paper the following notations are used. If $\Omega(x)=\sum_{d=1}^{D}\Omega_dx^d$ is a degree distribution with a maximum degree $D$, then $\Omega'(x)=\sum_{d=1}^{D}d\Omega_dx^{d-1}$ is the derivative of $\Omega(x)$ and $\Omega'(1)$ represents the average degree of $\Omega(x)$. We use a boldface letter to denote a vector and the $i^{th}$ entry of vector $\mbox{\textbf{V}}$ is denoted as $v_i$. We assume that all entries of vectors are chosen from the set $\{0,1\}$ unless otherwise specified, and $\mbox{n}(\textbf{V})=\sum_{i=1}^{\text{dim}(\textbf{V})}v_i$ represents the number of nonzero entries in vector $\mbox{\textbf{V}}$ of dimension $\text{dim}(\textbf{V})$. Also sets $\mathcal{A}_j^{(m)}$ and $\mathcal{A}^*(\textbf{V})$ are defined as follows:
\begin{gather}
\nonumber \mathcal{A}_j^{(m)}=\{\textbf{Y}|\text{dim}(\textbf{Y})=m, y_j=1\},\\
\nonumber \mathcal{A}^*(\textbf{V})=\{\textbf{Y}|\text{dim}(\textbf{Y})=\text{dim}(\textbf{V}), y_i=v_i, \forall v_i=1\},
\end{gather}
where $\mathcal{A}_j^{(m)}$ is the set of all vectors \textbf{Y} of dimension $m$ that their $j^{th}$ entries are $1$, and $\mathcal{A}^*(\textbf{V})$ is the set of all binary vectors \textbf{Y} which have the same dimension as \textbf{V} and for the positions of $v_i=0$, $y_i$ could be either 0 or 1; but for all positions of $v_i=1$, $y_i$ must be equal to 1 as well.

The rest of the paper is organized as follows. The system model is presented in Section II. In Section III and IV, we present the proposed FCC and PCC schemes, their designs, analysis and the degree distribution optimizations. Simulation and analytical results are shown in Section V. Finally, conclusions are drawn in Section VI.

\section{System model}
\label{sysmod}
\begin{figure}[t]
\centering
\includegraphics[scale=0.3]{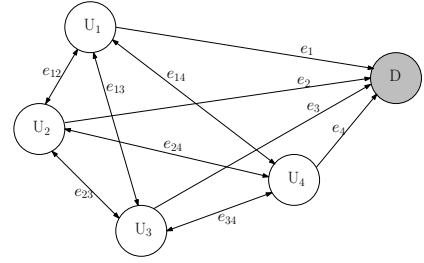}
\caption{M-user CMAC, $M=4$.}
\label{systemmodel4user}
\end{figure}

We consider a multi-user cooperative access network, shown in Fig. \ref{systemmodel4user}, where $M$ independent users ($\text{U}_1$, $\text{U}_2$,..., and $\text{U}_M$) in a network cooperate with each other to communicate with a common destination (D). Nodes in the network are assumed to be half-duplex, i.e. they cannot transmit and receive simultaneously. Each user first divide its message into $n$ packets with the same length of $T$. Then, by using a high rate LDPC code, $k$ encoded packet are generated, referred to as message symbols.  An LT code \cite{Luby} is then applied to encode the LDPC codeword at each user, generating potentially an infinite number of Raptor coded packets, referred to as coded symbols. In this paper, we define a symbol as a packet of length $T$, unless otherwise specified. For LT encoding, first an integer $d$ is obtained from a predefined degree distribution function $\Omega(x)=\sum_{d=1}^{D}\Omega_dx^d$, where $\Omega_d$ is the probability of the degree being $d$ and $D$ is the maximum degree. Then $d$ different message symbols are selected uniformly at random and added modulo 2 to generate one coded symbol. A time-division multiple-access scheme is then assumed for transmission, where each time frame (TF) is divided into $M$ equal length time slots (TS) and only one user transmits in each TS. More specifically, $\text{U}_i$ sends $N$ coded symbols in the $i^{th}$  TS of each TF.

In this paper, we consider a packet-erasure channel in cooperative systems. Let $e_{ij}$ and $e_{i}$ denote the packet erasure probability of the channel between $\text{U}_i$ and $\text{U}_j$ and that between $\text{U}_i$ and D, respectively, where $i,j=1,2,...,M$ and $i\ne j$. We assume that inter-user channels are reciprocal, i.e., $e_{ij}=e_{ji}$ ($i\ne j$). This model is appropriate for wireless networks where all information transmission is packetized and channel coding is used for each packet. More specifically, by considering non-decodable packets as an erasure, the channel can be treated as an erasure channel with a specific erasure probability. This model has been widely used in many existing work \cite{maymounkov2003rateless,byers2002informed,stefanovic2009raptor,dimakis2006distributed,lin2007data,aly2008raptor,cataldi2009corp}.

\section{Distributed Rateless Codes with Fully Coded Cooperation (FCC)}
\label{FCC}
In this section, we present the FCC scheme. We first introduce the FCC scheme for a two-user CMAC and then extend it to the general M-user case.
\subsection{FCC Scheme}
\label{fccscheme}
\begin{figure}[t]
\centering
\includegraphics[scale=0.38]{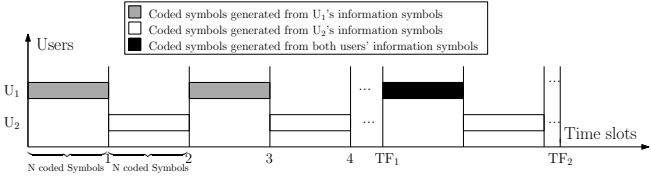}
\caption{The FCC scheme for a 2-user CMAC. $\text{TF}_{1}$ and $\text{TF}_2$ are the time slots that $\text{U}_1$ and $\text{U}_2$ start the cooperative phase, respectively. Time slots are of fixed duration, and in each time slot $N$ Raptor coded packets are transmitted.}
\label{schemesFCC}
\end{figure}

In the FCC scheme, each user's transmission is divided into two separate phases, a broadcast phase and a cooperative phase. In the broadcast phase, a user generates coded symbols from its own message symbols using degree distribution $\Phi^{(1)}(x)$ and transmits them to other users and the destination in its allocated TS in each TF. If information symbols of all users are successfully decoded at the destination before they are recovered by other users, the destination will send an acknowledgement to the users. Then, the users start transmitting new information symbols.  Otherwise, those users that have successfully recovered at least one other user's information symbols, start the cooperative phase while other users will still remain in the broadcast phase. In other words, users enter into the cooperation phase independently. In the cooperative phase, the user that has recovered information symbols of $m$ users, where $1\le m\le M-1$, generates coded symbols from message symbols of $m$ users and its own message symbols by using an LT code with a degree distribution $\Phi^{(m+1)}(x)$ and transmits them to the destination. When the destination completely decodes all users' information symbols, it sends an acknowledgment and all users start broadcasting new information symbols.
Fig. \ref{schemesFCC} shows the FCC scheme for a 2-user CMAC.  It is worth noting that in FCC, users need to perform both LT and LDPC decoding to recover other users' information symbols. We use the simple LT decoding algorithm \cite{Luby}, where in each iteration degree one coded symbols are found and then their connected message symbols are verified and removed from the bipartite graph. More details on the LT decoding process can be found in \cite{Fountain,Raptor,Luby}.
\subsection{Performance analysis of the FCC scheme}
\label{fccperf}
Let us first introduce the AND-OR tree analysis which has been extensively used to calculate the recovery probability of information symbols in rateless codes \cite{andor,UEP,DistUEP,AndOrDistLT,sejdinovic2009expanding}.

\begin{figure}[t]
\centering
\includegraphics[scale=0.4]{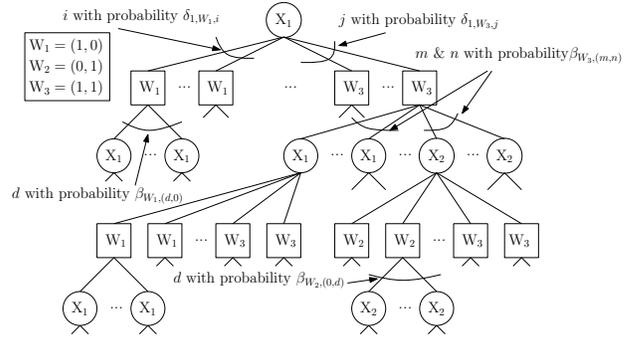}
\caption{AND-OR tree $T_{l,1}$, when $N_O=2$. $i$, $j$, $d$, $m$, and $n$ are variables which indicate the number of edges connected to the specified node.}
\label{andortree}
\end{figure}

Generally, an AND-OR tree  $T_{l,j}$ of depth $2l$ is defined as follows. The root of the tree is at depth 0 and its children are at depth 1. The children of nodes at depth $i$ are at depth $i+1$, where $1\le i\le2l-1$. The nodes at depth $0$, $2$,..., $2l$ are called OR nodes, and those at depth $1$, $3$,..., $2l-1$ are called AND nodes. We consider $N_O$ AND-OR trees, $T_{l,1}$, $T_{l,2}$,..., and $T_{l,N_O}$ with depth $2l$, and the root of $T_{l,j}$ is assumed to be a Type-$X_j$ OR-node ($j=1,2,..., N_O$). We also consider $2^{N_O}-1$ types of AND nodes, where each type is represented by a vector of dimension $N_O$, denoted by $\textbf{V}$. A Type-$\textbf{V}$ AND node has children of Type-$X_i$ OR nodes, if $v_i=1$. For example, Fig. \ref{andortree} shows an AND-OR tree, $T_{l,1}$, for the case of $N_O=2$. As shown in this figure, the root of $T_{l,1}$ is a Type-$X_1$ OR node and there are three types of AND nodes, namely Type-$\textbf{W}_1$, Type-$\textbf{W}_2$, and Type-$\textbf{W}_3$, where $\textbf{W}_1=(1,0)$, $\textbf{W}_2=(0,1)$, and $\textbf{W}_3=(1,1)$.

Let $\beta_{\textbf{V},\textbf{I}}$ denote the probability that a Type-$\textbf{V}$ AND node has $v_ji_j$ children of Type-$X_j$ OR nodes, where $j=1,2,...,N_O$, $\textbf{I}$ is a non-binary vector of dimension $N_O$. Let $\delta_{j,\textbf{V},i}$ denote the probability that a Type-$X_j$ OR node has  $i$ Type-$\textbf{V}$ AND children. Fig. \ref{andortree} shows these probabilities for the case of $N_O=2$. Each Type-$X_i$'s OR node at depth $2l$ are independently assigned a value of $0$ and $1$ with probabilities $p_{0,i}$ and $1-p_{0,i}$, respectively. Moreover, OR-nodes with no children are assigned a value of $0$ and AND-nodes with no children are assigned a value of $1$. By treating the trees as Boolean circuits, the following lemma gives the probability that the value of the root of a tree $T_{l,j}$ is zero.
\newtheorem{lemma}{Lemma}
\begin{lemma}[Generalized AND-OR tree lemma]
\label{PCCandortreelemma}
Let $p_{l,j}$ denote the probability that the value of the root of $T_{l,j}$ is zero. Then, $p_{l,j}$ can be calculated as follows.
\begingroup\makeatletter\def\f@size{9}\check@mathfonts
\def\maketag@@@#1{\hbox{\m@th\small\normalfont#1}}%
\begin{align}
\label{PCCandOrLemma}
\nonumber &p_{l,j}=\\
&\prod_{\textbf{V}\in \mathcal{A}_j^{(N_O)}}\delta_{j,\textbf{V}}\left(1-\sum_{d}\sum_{\textbf{I}\in \mathcal{I}_{\textbf{V},j,d}}\beta_{\textbf{V},\textbf{I}}\prod_{w=1}^{N_O}(1-p_{l-1,w})^{v_wi_w}\right),
\end{align}\endgroup
where $\mathcal{I}_{\textbf{V},j,d}=\{(v_1i_1,v_2i_2,...,v_{N_O}i_{N_O})|\sum_{m=1}^{N_O}v_mi_m=d, i_m\ge 1,\mbox{ for all}~ m\ne j\}$, and $\delta_{j,\textbf{V}}(x)=\sum_{i}\delta_{j,\textbf{V},i}x^i$.  %$d=\sum_{j=1}^{2L}v_ji_j$ and $||V||=\sum_{j=1}^{2L}v_j$.
\end{lemma}
The proof of this lemma is provided in Appendix \ref{PCCandortreelemmaProof}.
\begin{figure*}
\begingroup\makeatletter\def\f@size{9}\check@mathfonts
\def\maketag@@@#1{\hbox{\m@th\small\normalfont#1}}%
\begin{align}
\label{FCCandOrLemma}
p_{l,j}=\delta_{j,\textbf{W}_j}\left(1-\sum_{d=1}^{D-1}\beta_{\textbf{W}_j,d}(1-p_{l-1,j})^d\right)
\times\delta_{j,\textbf{W}_3}\left(1-\sum_{d=1}^{D-1}\sum_{i=1}^{d}\beta_{\textbf{W}_3,d-i,i}(1-p_{l-1,j})^{d-i}(1-p_{l-1,3-j})^{i}\right), ~l>1,
\end{align}\endgroup
\hrulefill
\end{figure*}
Such a generalization of AND-OR tree analysis is similar to the analysis of multi-edge type LDPC codes \cite{richardson2002multi}. However, the representation of AND-OR tree approach is clearer and more convenient for analysis as it perfectly matches to the decoding process of LT codes. As shown in \cite{andor,OnLineCodes}, in LT decoding \cite{Luby}, variable nodes actually perform the logical OR operation and check nodes perform the logical AND operation. This is exactly the same to what happens in AND-OR trees. More details of AND-OR tree analytical approach for different types of rateless codes can be found in \cite{UEP,sejdinovic2009expanding}.

\subsubsection{AND-OR Tree Analysis of the FCC Scheme for a 2-user CMAC}
\begin{figure}[t]
\centering
\includegraphics[scale=0.3]{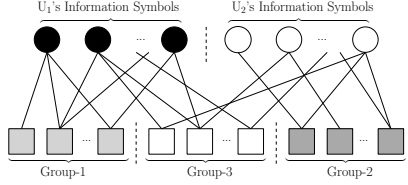}
\caption{Three different groups of coded symbols that have been received at the destination.}
\label{groupsFig}
\end{figure}

Since coded packets received at the destination are generated from message symbols of both users, three groups of Raptor coded packets exist at the destination (Fig. \ref{groupsFig}). In the bipartite graph, coded symbols that are only connected to $\text{U}_1$ and $\text{U}_2$'s message symbols are referred to as Group-1 and Group-2 coded symbols, respectively, and coded symbols that are  connected to message symbols of both users are called Group-3 coded symbols.

In FCC,  message symbols of $\text{U}_1$ and $\text{U}_2$ are mapped to Type-$X_1$ and Type-$X_2$ OR nodes. Furthermore, three groups of coded symbols are mapped to three different types of AND nodes, namely Type-$\textbf{W}_1$, Type-$\textbf{W}_2$, and Type-$\textbf{W}_3$ AND nodes. By using Lemma \ref{PCCandortreelemma}, the probability that a message symbol of $\text{U}_j$ is not recovered after $l$ iterations, $p_{l,j}$,  is given by (\ref{FCCandOrLemma}), where, $p_{0,j}=1$, and $D$ is the maximum degree. Other parameters have been shown in Fig. \ref{andortree} for clarity. To compute the error probability of each user's message at the destination by using (\ref{FCCandOrLemma}), we only need to compute probabilities $\beta_{\textbf{W}_j,d}, \beta_{\textbf{W}_3,d-i,i}$, functions $\delta_{j,\textbf{W}_j}(x)$, and $\delta_{j,\textbf{W}_3}(x)$, which are given in the following lemma.
\begin{lemma}
\label{ANDORlemmaFCC2users}
We assume that in the FCC scheme, the destination receives $N_1$ and $N_2$ coded symbols from $\text{U}_1$ and $\text{U}_2$ in the broadcast phase, respectively, and $N_3$ coded symbols in the cooperative phase from both users. If users use degree
distributions $\Phi^{(1)}(x)$ and $\Phi^{(2)}(x)$ in the broadcast and cooperative phase, respectively, then we have
\begingroup\makeatletter\def\f@size{9}\check@mathfonts
\def\maketag@@@#1{\hbox{\m@th\small\normalfont#1}}%
\begin{multline}
\beta_{\textbf{W}_j,d}=\frac{N_j}{N_j+N_3\Phi^{(2)}(0.5)}\frac{(d+1)\Phi^{(1)}_{d+1}}{\mu^{(1)}}\\
+\frac{N_3\Phi^{(2)}(0.5)}{N_j+N_3\Phi^{(2)}(0.5)}\frac{(d+1)\Phi^{(2)}_{d+1}}{\mu^{(2)}},
\end{multline}
\begin{align}
&\beta_{\textbf{W}_3,i,j}=\frac{(i+j+1)\Phi^{(2)}_{i+j+1}}{\mu^{(2)}}\frac{\dbinom{k}{i}\dbinom{k}{j}}{\dbinom{2k}{i+j+1}},\\
&\delta_{j,\textbf{W}_j}(x)=e^{\alpha_{j}(x-1)},\alpha_{j}=\frac{N_j\mu^{(1)}+N_3\Phi^{(2)}(0.5)\mu^{(2)}}{k},\\
&\delta_{j,\textbf{W}_3}(x)=e^{\alpha_3(x-1)},\alpha_3=\frac{N_3(1-2\Phi^{(2)}(0.5))\mu^{(2)}}{2k},
\end{align}\endgroup
where $\mu^{(1)}=\Phi^{(1)'}(1)$ and $\mu^{(2)}=\Phi^{(2)'}(1)$.
\end{lemma}
Eq. (4) and (5) are arisen from the fact that a Type-$\textbf{W}_j$ AND node can be either received in the broadcast phase or cooperative phase. This means that the number of coded symbols that have been  received in the cooperative phase and are only connected to $U_j$ is $N_3\Phi^{(2)}(0.5)$. Equations (6)-(7) are also obtained from the fact that in LT codes, the variable node's degree follows the Poisson distribution with parameter $\alpha$ \cite{Raptor}, where $\alpha$ is the average degree of variable nodes.
\subsubsection{AND-OR Tree Analysis of the FCC Scheme for the General M-user CMAC}
For the general M-user case, we assume that each user will use $\Phi^{(n)}(x)$ as its degree distribution when it has recovered $n-1$ users' messages ($1\le n\le M$). Since we have $M$ different users, we have $M$ different types of message symbols; thus we consider $M$ AND-OR trees, where the root of $T_{l,j}$ is a Type-$X_j$ OR node ($j=1,2,...,M$). Each type of OR-nodes represents one user's message symbols, and each type of AND-nodes represents coded symbols, which are generated from message symbols of a set of users. Let $N_{\textbf{V}}$ denote the total number of coded symbols, which are generated from message symbols of the users in set $\textbf{V}$. By using Lemma \ref{PCCandortreelemma} and setting $N_O=M$ and $p_{0,j}=1$, (\ref{PCCandOrLemma}) will give the probability that a message symbol of $\text{U}_j$ is not recovered after $l$ iterations. To complete the analysis we only need to calculate $\beta_{\textbf{V},\textbf{I}}$ and $\delta_{j,\textbf{V}}$, which will be given in the following lemma.
\begin{lemma}
\label{FCCMuserAndOrtree}
$\beta_{\textbf{V},\textbf{I}}$ and $\delta_{j,\textbf{V}}$ are calculated as follows:
\begingroup\makeatletter\def\f@size{9}\check@mathfonts
\def\maketag@@@#1{\hbox{\m@th\small\normalfont#1}}%
\begin{align}
\nonumber \beta_{\textbf{V},\textbf{I}}=
&\sum_{\textbf{Y}\in\mathcal{A}^{*}(\textbf{V})}\frac{N_{\textbf{Y}}\Phi^{*(\text{n}(\textbf{Y}))}(\frac{\text{n}(\textbf{V})}{\text{n}(\textbf{Y})})(\frac{1}{\text{n}(\textbf{Y})})^{\text{n}(\textbf{V})}(1+d)\Phi^{(\text{n}(\textbf{Y}))}_{d+1}}{T_\textbf{V}\mu^{(\text{n}(\textbf{Y}))}}\\
\label{betaFCCV}
&~~~~~~~~~\times\frac{\prod_{j=1}^{M}\dbinom{k}{v_ji_j}}{\dbinom{\text{n}(\textbf{V})k}{d+1}},
\end{align}
\begin{align}
\label{deltaFCCMV} &\delta_{j,\textbf{V}}(x)=e^{\alpha_{j,\textbf{V}}(x-1)},\\
\nonumber&\alpha_{j,\textbf{V}}=\frac{\sum_{\textbf{V}\in \mathcal{A}^{*}(\textbf{V})}N_\textbf{Y}\Phi^{*(\text{n}(\textbf{Y}))}(\frac{\text{n}(\textbf{V})}{\text{n}(\textbf{Y})})(\frac{1}{\text{n}(\textbf{Y})})^{\text{n}(\textbf{V})}\mu^{(\text{n}(\textbf{Y}))}}{\text{n}(\textbf{V})k},
\end{align}\endgroup
where $\Phi^{*(n)}(x)=\sum_{d=n}^{D}\Phi^{(n)}_dx^{d-n}$, $\mu^{(n)}=\Phi^{(n)'}(1)$, $N_\textbf{Y}$ is the number of coded symbols generated from users in set $\textbf{Y}$, and $T_\textbf{V}$ is the total number of Type-$\textbf{V}$ AND nodes given by
\begingroup\makeatletter\def\f@size{9}\check@mathfonts
\def\maketag@@@#1{\hbox{\m@th\small\normalfont#1}}%
\begin{align}
\label{NumberofV}
T_\textbf{V}=\sum_{\textbf{Y}\in \mathcal{A}^{*}(\textbf{V})}N_{\textbf{Y}}\Phi^{*(\text{n}(\textbf{Y}))}(\frac{\text{n}(\textbf{V})}{\text{n}(\textbf{Y})})(\frac{1}{\text{n}(\textbf{Y})})^{\text{n}(\textbf{V})}.
\end{align}\endgroup
\end{lemma}
The proof of this lemma is provided in Appendix \ref{FCCMuserAndOrtreeProof}.
\subsection{Degree Distribution Optimization of FCC}
\label{fccdeg}
From Lemma \ref{FCCMuserAndOrtree}, we can define an optimization problem to minimize the error probability
of $\text{U}_j$ at iteration $l$, denoted by $p_{l,j}$'s. However, to find the optimum degree distributions, $N_{\textbf{V}}$ values need to be known beforehand. But, $N_{\textbf{V}}$ values cannot be computed without the knowledge of the channel erasure probability. Moreover, the degree distribution needs to be independent of the channel erasure probability in order to have a reasonable performance at all channel conditions. We propose a suboptimum solution for the degree distribution, and later we will show that the obtained degree distributions perform well in all channel conditions.

Let us consider that the destination already knows the message symbols of $m$ users; therefore, all edges connected to these symbols can be removed from the bipartite graph at the destination. So the coded symbols have the following degree distribution
\begingroup\makeatletter\def\f@size{9}\check@mathfonts
\def\maketag@@@#1{\hbox{\m@th\small\normalfont#1}}%
\begin{align}
\label{FCCSubDeg}
\Phi_d^{(M,m)}=\sum_{w=0}^{mk}\Phi_{d+w}^{(M)} \frac{\dbinom{mk}{w}\dbinom{(M-m)k}{d}}{\dbinom{Mk}{d+w}},
\end{align}\endgroup
where $m=1,2,...,M-1$. In fact, a degree-$d$ coded symbol is actually of degree $d+w$, and connected to $w$ known message symbols and $d$ unknown message symbols, before removing all edges connected to the known message symbols. This occurs with probability of $\dbinom{mk}{w}\dbinom{(M-m)k}{d}/\dbinom{Mk}{d+w}~$ due to the fact that message symbols are selected uniformly at random from $Mk$ message symbols, and $mk$ of them are known at the destination. Since $w$ varies from $0$ to $mk$, then (\ref{FCCSubDeg}) is self-evident.

Similar to the optimization problem in \cite{Raptor}, for given $m$, $\epsilon_m$ and $\delta$, to guarantee that a fraction $1-\delta$ of the $M-m$ users' message symbols can be recovered with a high probability at a minimum overhead, $\Phi^{(M,m)}(x)$ should satisfy the following condition \cite{CoopLowMACtrans}
\begingroup\makeatletter\def\f@size{9}\check@mathfonts
\def\maketag@@@#1{\hbox{\m@th\small\normalfont#1}}%
\begin{gather}
\label{OptFCCProb}
\Phi^{'(M,m)}(x)\ge \frac{-\text{ln}\left(1-x-c\sqrt{\frac{(1-x)}{(M-m)k}}\right)}{1+\epsilon_m},
\end{gather}\endgroup
where $x\in[0,1-\delta]$. This condition ensures us that all message symbols can be decoded at a minimum overhead by keeping the number of degree one coded symbols larger than or equal to $c\sqrt{(1-x)k}$ for some constant $c$ at each iteration. Furthermore, in (\ref{OptFCCProb}), $\epsilon_m$ can be seen as a parameter that should be found in the optimization algorithm. Let $r_m=\frac{1}{1+\epsilon_m}$, then we need to find $\Phi^{(M)}(x)$ and $r_m$ ($m=0,1,...,M-1$) that satisfy (\ref{OptFCCProb}). For simplicity reasons, we consider a linear objective function, $\sum_{m=0}^{M-1}r_m$. This approach was originally proposed in \cite{CoopLowMACtrans} based on finding induced degree distribution when a number of users' messages are known at the destination. Following the similar approach as in \cite{CoopLowMACtrans}, the optimization problem can be formulated as follows
\begingroup\makeatletter\def\f@size{9}\check@mathfonts
\def\maketag@@@#1{\hbox{\m@th\small\normalfont#1}}%
\begin{align}
\nonumber\text{maximize} \sum_{m=0}^{M-1}r_m
\end{align}\endgroup
subject to
\begingroup\makeatletter\def\f@size{9}\check@mathfonts
\def\maketag@@@#1{\hbox{\m@th\small\normalfont#1}}%
\begin{eqnarray}
\label{FCCwholeCondition}
(i).
&\nonumber\sum_{d=1}^{D}\sum_{w=0}^{mk}d{}x^{d-1}{}\Phi_{d+w}^{(M)}\frac{\dbinom{mk}{w}\dbinom{(M-m)k}{d}}{\dbinom{Mk}{d+w}}\\
\nonumber&\ge-r_m\mbox{ln}\left(1-x-c\sqrt{\frac{(1-x)}{(M-m)k}}\right), 0\le m\le M-1\\
\nonumber(ii).&\displaystyle\sum_{i}\Phi^{(M)}_i=1 \\
\nonumber(iii).&\displaystyle0\le\Phi^{(M)}_i\le1,~~~~0\le i \le k
\label{sumCond}
\end{eqnarray}\endgroup
where, constraint ($i$) is obtained by substituting (\ref{FCCSubDeg}) in (\ref{OptFCCProb}) and $x\in[0,1-\delta]$. Note that constraint ($i$) is actually $M$ constraint ($m$ varies between 0 and $M-1$) over $x\in[0,1-\delta]$. Moreover, constraints ($i$)-($iii$) and the objective function are all linear in terms of $\Phi^{(M)}(x)$; thus, the optimization problem can be solved by a linear programming approach. Table \ref{OptDegFCC} lists some optimized degree distributions for different $M$ values, which are found by using a linear programming optimization approach.

\begin{table}[t]
\caption{Degree Distributions for various values of $M$ when $k=10000$ and $\delta=0.01$; $\mu$ is the average degree of a coded symbol}
\label{OptDegFCC}
\centering
\scriptsize
\begin{tabular}{|c||c|c|c|c|}
\hline
$M$&1&2&3&4\\
\hline
\hline
$\Omega_1$&0.0098&0.0067&0.0050&0.0061\\
\hline
$\Omega_2$&0.4949&0.4749&0.4446&0.4243\\
\hline
$\Omega_3$&0.1597&0.1543&0.1050&0.1843\\
\hline
$\Omega_4$&0.1095&0.0884&0.1691&0.0714\\
\hline
$\Omega_5$&&0.0550&&\\
\hline
$\Omega_6$&0.0437&&&\\
\hline
$\Omega_7$&0.0774&&&\\
\hline
$\Omega_8$&&0.0966&&\\
\hline
$\Omega_9$&&&&0.2249\\
\hline
$\Omega_{11}$&&&0.1753&\\
\hline
$\Omega_{14}$&0.0026&&&\\
\hline
$\Omega_{15}$&0.0661&&&\\
\hline
$\Omega_{20}$&&0.0466&&\\
\hline
$\Omega_{21}$&&0.0184&&\\
\hline
$\Omega_{50}$&0.0358&0.0586&0.1007&0.0887\\
\hline
\hline
$\mu$&5.5442&7.0752&8.8531&8.15\\
\hline
\end{tabular}
\end{table}

\section{Distributed Rateless Codes with Partially Coded Cooperation (PCC)}
\label{PCC}
In this section, we will design robust rateless codes with enhanced partial decoding performance for the PCC scheme, fully exploiting the coded cooperation to improve the overall system performance.

\subsection{PCC Scheme}
In the PCC scheme, in the first TF ($\text{TF}_1$), each user generates $N$ coded symbols from its own message symbols using an LT code with a degree distribution $\Omega(x)$, and broadcasts them to all other users and the destination in its allocated TS. Upon receiving coded symbols in $\text{TF}_1$, each user starts the decoding process (LT decoding) to recover the message symbols of other users. We assume that $\text{U}_i$ has recovered $s_{i,j}^{(l)}$ message symbols from $\text{U}_j$ in $\text{TF}_l$, where $0\le s_{i,j}^{(l)}\le k$ and $i\ne j$. In $\text{TF}_{2}$, $\text{U}_i$ generates $N$ coded symbols by using its $k$ message symbols as well as message symbols of other users that have been decoded in the previous TF in $\text{U}_i$, and broadcasts them. Generally, in each TF, each user partially decodes other users' message symbols from the received coded symbols in all previous TFs, and generates coded symbols using its own message symbols and recovered message symbols of other users. When the destination receives an adequate number of coded symbols from users and completely decodes all users' information symbols, it sends an acknowledgment to all users. Each user will then start sending new information symbols.

\begin{figure}[t]
\centering
\includegraphics[scale=0.4]{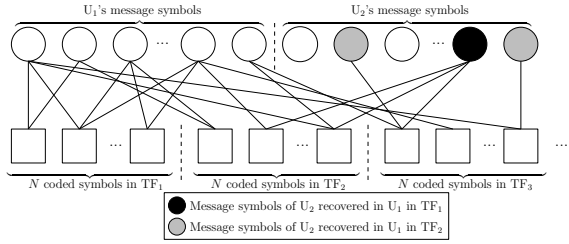}
\caption{The PCC scheme in $\text{U}_1$ for a 2-user CMAC.}
\label{schemesPCC}
\end{figure}

Fig. \ref{schemesPCC} shows the PCC scheme for a 2-user CMAC. As can be seen in this figure, $\text{U}_1$ generates coded symbols from its own message symbols in $\text{TF}_1$. In $\text{TF}_2$, it generates coded symbols from its own message symbols and those from $\text{U}_2$, which have been decoded in $\text{U}_1$ in $\text{TF}_1$, and so on.  Note that the PCC scheme does not require the complete decoding of other users' messages at each user before starting the cooperation process. Cooperation starts as soon as the users are capable of recovering at least one message symbol from other users. Note that in the proposed partially coded cooperation scheme, users need to perform only LT decoding, which leads to a lower decoding complexity at users compared to the FCC scheme.
\subsection{Practical Considerations in PCC}
Due to the random nature of the erasure channel, each user can possibly decode any subset of the entire block of data of other users in one TF. This means that in the next TF, other users need to know exactly which subset has been decoded at other users in the previous TF in order to perform the encoding and decoding process. That is, in each TF, $\text{U}_i$ needs to inform $\text{U}_j$ that which information packets of $\text{U}_j$ have been decoded at $\text{U}_i$ in the previous TF, where $i\ne j$. To address this issue, we can simply attach a header to the entire block of data that each user sends in each TF and indicate which information packet of other users have been decoded, thus users can partially decode information packets of other users accordingly. This requires at most $(M-1)k$ bits in the header, where each control overhead bit indicates whether a specific message packet out of $k$ original packets has been decoded at the other user or not. The ratio of the control overhead and the overall message length will then be at most $\frac{(M-1)k}{TN}$, which is independent of the channel erasure probabilities. In practice, $T$ is about 512 to 1024 bytes, so the control overhead is about 2\% to 1\%, when $k/N=10$ and $M=2$. To further reduce the control overhead, we can perform the decoding process at each user in every $F$ time frames. This means that each user tries to decode other users' messages after $F$ time frames, and so it uses the decoded symbols to generate coded symbols for the next $F$ time frames. This approach will obviously lead to some performance loss, but it decreases the control overhead as well as the decoding complexity at users by a factor $F$.

Table \ref{ControlOverhead} shows the control overhead and performance loss for PCC with different values of $F$, when $e_1=0.2$ and $e_2=0.6$. In this table the control overhead has been shown in percent and the performance loss is shown in dB. The performance loss was calculated by comparing the average system throughout when $F\ne1$ with the case of $F=1$ and averaging the loss over all inter-user channel erasure probabilities. As can be seen in this table, by increasing $F$, we can achieve a considerably lower control overhead with little performance losses. For example, when $F=3$ the required control overhead is only 0.32\% and the performance loss is 0.36 dB, which is negligible.
\begin{table}[t]
\caption{Control overhead (C.O.) and average performance loss (P.L) for the 2-user CMAC, when $e_1=0.2$, $e_2=0.6$, and $T=1024$.}
\label{ControlOverhead}
\centering
\scriptsize
\begin{tabular}{|c||c|c||c|c|}
\hline
&\multicolumn{2}{|c||}{$k/N=10$}&\multicolumn{2}{|c||}{$k/N=5$}\\
\hline
F&C.O.\%&P.L. [dB]&C.O.\%&P.L. [dB]\\
\hline
\hline
1&0.97&0&0.48&0\\
\hline
2&0.48&0.23&0.24&0.31\\
\hline
3&0.32&0.36&0.17&0.48\\
\hline
4&0.24&0.62&0.12&0.89\\
\hline
\end{tabular}
\end{table}
\normalsize

\subsection{Performance Analysis of the PCC Scheme}
In this section, we analyze the PCC scheme using AND-OR tree analysis by first, considering a two-user CMAC and then extending it to the general M-user case. % The recovery rate at users and the destination will also be analyzed.
\subsubsection{Code Performance at Users in a 2-user CMAC}
Let $s_{2,1}^{(i)}$ and $s_{1,2}^{(i)}$ represent the total numbers of the decoded message symbols of $\text{U}_1$ at $\text{U}_2$ and that of $\text{U}_2$ at $\text{U}_1$, respectively, in $\text{TF}_i$, and $e_{12}=e_{21}=e$. Since in $\text{TF}_i$, $s_{2,1}^{(i)}$ message symbols of $\text{U}_1$ and $s_{1,2}^{(i)}$ message symbols of $\text{U}_2$ are known at $\text{U}_2$ and $\text{U}_1$, respectively, $\text{U}_1$ and $\text{U}_2$ can remove all edges connected to the known message symbols in their associated bipartite graphs. Similar to (\ref{FCCSubDeg}), the resulting degree distribution of coded symbols at $\text{U}_2$ can be calculated as follows:
\begingroup\makeatletter\def\f@size{9}\check@mathfonts
\def\maketag@@@#1{\hbox{\m@th\small\normalfont#1}}%
\begin{align}
\label{OutDegPCC}
\Omega^{(i+1)}_{2,d}=\sum_{l=0}^{s_{1,2}^{(i)}} \Omega_{d+l} \frac{\dbinom{s_{1,2}^{(i)}}{l}\dbinom{k}{d}}{\dbinom{k+s_{1,2}^{(i)}}{d+l}},~~~ d=0,1,...,k.
\end{align}
\endgroup
In each TF, $\text{U}_2$ actually receives $N(1-e)$ coded symbols on average from $\text{U}_1$. Each user tries to decode the other user's message from the received coded symbols in $\text{TF}_{i+1}$ and previous TFs, so the average number of coded symbols received at $\text{U}_2$ in $\text{TF}_{i+1}$ will be $N^{(i+1)}_{2}=(i+1)N(1-e)$.

Therefore, the probability that a coded symbol is of degree $d$ in $\text{TF}_{i+1}$ is
\begingroup\makeatletter\def\f@size{9}\check@mathfonts
\def\maketag@@@#1{\hbox{\m@th\small\normalfont#1}}%
\begin{align}
\label{PCCOutDegFinal}
\Delta^{(i+1)}_{2,d}=\frac{1}{i+1}\sum_{j=1}^{i+1}\Omega^{(j)}_{2,d},~~~d=0,1,2,...,k
\end{align}
\endgroup
which arises from the fact that a specific coded symbol is received in $\text{TF}_{j}$ with the probability of $\frac{N(1-e)}{N^{(i+1)}_{2}}$ and it is of degree $d$ with the probability of $\Omega^{(j)}_{2,d}$. %So we have

\begin{figure}[t]
\centering
\includegraphics[width=8.5cm, height=5.5cm]{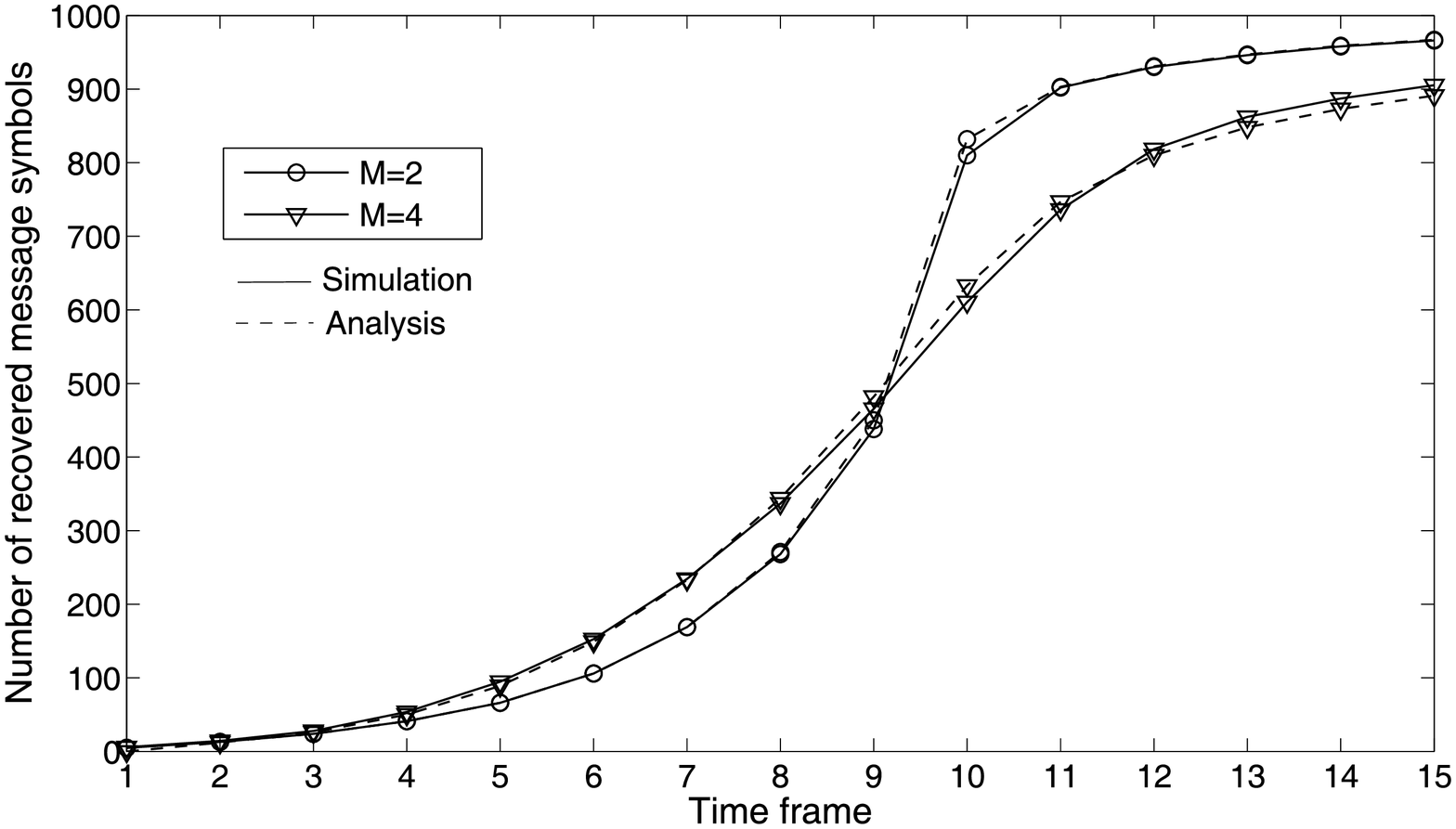}
\caption{The average number of recovered message symbols in each time frame at each user, when the erasure probability of inter-user channels are 0. }
\label{FigRatioOfRec2}
\end{figure}

In the PCC scheme, the degree distribution of coded symbols in $\text{TF}_{i+1}$ at $\text{U}_2$ is $\Delta^{(i+1)}_{2}(x)=\sum_{d=1}^{k}\Delta^{(i+1)}_{2,d}x^d$. Similar to \cite{Raptor} and by following a similar approach as for Lemma 3 in \cite{UEP}, the probability that a message symbol of $\text{U}_1$ is not recovered at $\text{U}_2$ after $l$ iterations, can be calculated as follows.
\begingroup\makeatletter\def\f@size{9}\check@mathfonts
\def\maketag@@@#1{\hbox{\m@th\small\normalfont#1}}%
\begin{align}
\label{AndOrPCC}
p_{2,l}^{(i+1)}=e^{(-\alpha^{(i+1)}\delta^{(i+1)}_{2}(1-p_{2,l-1}^{(i+1)}))}, l>1,
\end{align}\endgroup
where $p_{2,0}=1$, $\delta^{(i+1)}_{2}(x)=\Delta^{(i+1)'}_{2}(x)/\Delta^{(i+1)'}_{2}(1)$ and $\alpha^{(i+1)}_{2}=\frac{N_{2}^{(i+1)}}{k}\Delta_{2}^{(i+1)'}(1)$.
We can also easily prove that $p_{2,l}^{(i+1)}$ has the following properties.
\begin{lemma}
\label{pldecreaseLI}
$p_{2,l}^{(i+1)}$ has the following properties:

$(1)$. $p_{2,l}^{(i+1)}$ is a decreasing function of the number of iterations $l$.

$(2)$. $p_{2,l}^{(i+1)}$ is a decreasing function of the number of TFs $i$.
\end{lemma}
The proof of this lemma is provided in Appendix \ref{pldecreaseLIProof}. As $k$ and $N$ go to infinity, the number of recovered message symbols in $\text{TF}_{(i+1)}$, $s_{2,1}^{(i+1)}$, can be approximated by  $k(1-p_{2,l}^{(i+1)})$ when $l$ goes to infinity. In Lemma \ref{pldecreaseLI}, we have shown that $p_{2,l}^{(i)}$ decreases as $i$ increases, so $s_{2,1}^{(i+1)}$ increases with $i$. Thus, the number of message symbols recovered at each user increases with the number of TFs.  Fig. \ref{FigRatioOfRec2} shows the average number of recovered message symbols in each TF at each user when $k=1000$, $N=100$, and $\Omega(x)=0.05x+0.55x^2+0.25x^4+0.05x^6+0.1x^8$. Analytical results have also shown an excellent agreement with the simulation results. It is worth noting that due to symmetry, the average number of message symbols recovered at each user is the same as for the other user, i.e., $s_{2,1}^{(i+1)}=s_{1,2}^{(i+1)}$.

\subsubsection{Code Performance at the Destination in a 2-user CMAC}
Without loss of generality, we divide $\text{U}_j$'s message into $L_j$ parts and assume that the $i^{th}$ part is the fraction of its message recovered by the other user in $\text{TF}_i$. Specifically, we divide  $\text{U}_1$'s and $\text{U}_2$'s messages into $L_1$ and $L_2$ different parts, $P^{(1)}_1$, $P^{(1)}_2$, ..., $P^{(1)}_{L_1}$, and $P^{(2)}_1$, $P^{(2)}_2$, ..., $P^{(2)}_{L_2}$, respectively, where $P^{(1)}_i$ and $P^{(2)}_i$ are parts of $\text{U}_1$'s and $\text{U}_2$'s messages decoded in $\text{TF}_i$. The length of $P^{(1)}_i$ and $P^{(2)}_i$ are considered to be $p^{(1)}_i$ and $p^{(2)}_i$, respectively, where the length of different parts are not necessarily the same.

Next, let us use the result of Lemma \ref{PCCandortreelemma} to analyze the error performance of the PCC scheme at the destination. The set of message symbols $P^{(1)}_1$, $P^{(1)}_2$, ..., $P^{(1)}_{L_1}$ are mapped to Type-$X_1$, Type-$X_2$,..., and Type-$X_{L_1}$ OR nodes, and $P^{(2)}_1$, $P^{(2)}_2$, ..., $P^{(2)}_{L_1}$ are mapped to Type-$X_{L_1+1}$, Type-$X_{L_1+2}$,..., and Type-$X_{L_1+L_2}$ OR nodes, respectively. Since in each TF, coded symbols are generated from parts of both users' messages, we can consider different types of coded symbols. Specifically, we consider $2^L-1$ ($L=L_1+L_2$) different types of AND nodes due to the fact that each coded symbol can be generated from different sets of users' message parts. Each type of AND nodes is represented by a vector of dimension $L$, and its $i^{th}$ entry is set to 1 if it has at least one Type-$X_i$ AND node as its children; otherwise, it is set to 0.
\begin{lemma}
\label{PCClemmaFinal}
Let $p_{l,j}$ denote the probability that a message symbol in $P^{(1)}_j$ is not recovered after $l$ iterations.  If $S_\textbf{V}=\sum_{t=1}^{M}\sum_{i=1}^{L_t}v^{(t)}_ip^{(t)}_i$, where $M=2$, $v^{(t)}_i=v_g$, and $g=i+\sum_{a=1}^{t-1}L_t$, then $p_{l,j}$ can be calculated as follows.
\begingroup\makeatletter\def\f@size{9}\check@mathfonts
\def\maketag@@@#1{\hbox{\m@th\small\normalfont#1}}%
\begin{align}
\nonumber &p_{l,j}=\\
&\prod_{\textbf{V}\in \mathcal{A}^{(L)}_j}\delta_{j,\textbf{V}}\left(1-\sum_{d=0}^{D-1}\sum_{\textbf{I}\in \mathcal{I}_{\textbf{V},j,d}}\beta_{\textbf{V},\textbf{I}}\prod_{w=1}^{L}(1-p_{l-1,w})^{v_wi_w}\right),
\label{PCCLemma}
\end{align}\endgroup
where $p_{0,j}=1$ and
\begingroup\makeatletter\def\f@size{9}\check@mathfonts
\def\maketag@@@#1{\hbox{\m@th\small\normalfont#1}}%
\begin{align}
\label{betaPCC}
&\beta_{\textbf{V},\textbf{I}}=\frac{(1+d)\Omega_{d+1}}{\mu}\frac{\displaystyle\prod_{t=1}^{M}\prod_{j=1}^{L_t}\dbinom{p^{(t)}_j}{v^{(t)}_ji^{(t)}_j}}{\dbinom{S_\textbf{V}}{d+1}},\\
\label{deltaPCC}
&\delta_{j,\textbf{V}}(x)=e^{\alpha_{j,\textbf{V}}(x-1)}, ~~\alpha_{j,\textbf{V}}=\frac{\mu T_\textbf{V}}{S_\textbf{V}},
\end{align}\endgroup
where $T_\textbf{V}$ is the number of Type-$\textbf{V}$ AND nodes.
\end{lemma}
\begin{IEEEproof}
Using Lemma \ref{FCCMuserAndOrtree} and replacing $\Phi^{(n)}(x)$ by $\Omega(x)$ for all values of $n$ and considering that the message length of $P^{(i)}_j$ is $p^{(i)}_j$, (\ref{betaPCC}) and (\ref{deltaPCC}) are easily obtained from (\ref{betaFCCV}) and (\ref{deltaFCCMV}), respectively.
\end{IEEEproof}
\subsubsection{Code Performance at Users in a General M-user CMAC}
We assume that all inter-user channels have the same erasure probability $e$. Since each user receives on average $N(1-e)$ coded symbols from each of other users in each TF, after removing all edges connected to the known message symbols from the bipartite graph in $\text{U}_1$,  the degree distribution of coded symbols received at $\text{U}_1$, which have been transmitted from $\text{U}_w$ is as follows
\begingroup\makeatletter\def\f@size{9}\check@mathfonts
\def\maketag@@@#1{\hbox{\m@th\small\normalfont#1}}%
\begin{align}
\label{OutputdegreeMuser2}
\Omega_{1,w,d}^{(i+1)}=\sum_{l=0}^{s_{1,w}^{(i)}}\Omega_{d+l} \frac{\dbinom{\sum_{j=2}^{M}s_{w,j}^{(i)}}{l}\dbinom{k}{d}}{\dbinom{k+\sum_{j=2}^{M}s_{w,j}^{(i)}}{d+l}},  ~d=0,1,...,k.
\end{align}\endgroup
Due to the symmetry and assumption that all inter-user channels have the same erasure probability, the average number of recovered message symbols from each user at other users are the same. Let $s^{(i)}$ denote the average number of message symbols of each user recovered by each other user in $TF_i$; thus, $s_{j,w}^{(i)}=s^{(i)}$ for all $j\ne w$. Therefore, (\ref{OutputdegreeMuser2}) becomes
\begingroup\makeatletter\def\f@size{9}\check@mathfonts
\def\maketag@@@#1{\hbox{\m@th\small\normalfont#1}}%
\begin{align}
\label{NewDegreeSymmetry}
\Omega_{1,w,d}^{(i+1)}=\sum_{l=0}^{s^{(i)}}\Omega_{d+l} \frac{\dbinom{(M-1)s^{(i)}}{l}\dbinom{k}{d}}{\dbinom{k+(M-1)s^{(i)}}{d+l}}, ~d=0,1,...,k.
\end{align}\endgroup
Since in $\text{TF}_{i+1}$, $\text{U}_1$ receives the average number of $N(1-e)$ coded symbols from each user, and according to (\ref{NewDegreeSymmetry}), which does not depend on $w$, the degree distribution of all received coded symbols in $\text{TF}_{i+1}$ is the same as (\ref{NewDegreeSymmetry}). So the degree distribution of the received coded symbols at $\text{U}_1$ in $\text{TF}_{i+1}$ is as follows
\begingroup\makeatletter\def\f@size{9}\check@mathfonts
\def\maketag@@@#1{\hbox{\m@th\small\normalfont#1}}%
\begin{align}
\label{MuserTFdegree}
\Omega_{1,d}^{(i+1)}=\sum_{l=0}^{s^{(i)}}\Omega_{d+l} \frac{\dbinom{(M-1)s^{(i)}}{l}\dbinom{k}{d}}{\dbinom{k+(M-1)s^{(i)}}{d+l}},  ~d=0,1,...,k.
\end{align}\endgroup
Similar to the 2-user case, the total number of received coded symbols at $\text{U}_1$ from all other users till $\text{TF}_{i+1}$ is given by $N_1^{(i+1)}=\sum_{j=1}^{i+1}(M-1)N(1-e)$,
and the degree distribution of the total coded symbols received at $\text{U}_1$ will be
\begingroup\makeatletter\def\f@size{9}\check@mathfonts
\def\maketag@@@#1{\hbox{\m@th\small\normalfont#1}}%
\begin{align}
\label{MusserDeltafinal}
\Delta^{(i+1)}_{1,d}=\frac{1}{i+1}\sum_{j=1}^{i+1}\Omega^{(j)}_{1,d},~~~d=0,1,2,...,k.
\end{align}\endgroup
Similar to (\ref{AndOrPCC}), the probability that a message symbol of another user is not recovered at $U_1$ in $TF_{i+1}$ can be calculated as follows.
\begingroup\makeatletter\def\f@size{9}\check@mathfonts
\def\maketag@@@#1{\hbox{\m@th\small\normalfont#1}}%
\begin{align}
\label{OptDegrOpt}
p_{1,l}^{(i+1)}=e^{(-\alpha^{(i+1)}_1\delta^{(i+1)}_{1}(1-p_{1,l-1}^{(i+1)}))}, l>1,
\end{align}\endgroup
where $p_{1,0}=1$, $\delta^{(i+1)}_{1}(x)=\Delta^{(i+1)'}_{1}(x)/\Delta^{(i+1)'}_{1}(1)$ and $\alpha^{(i+1)}_{1}=\frac{N_{1}^{(i+1)}}{(M-1)k}\Delta_{1}^{(i+1)'}(1)$. When $k$ and $N$ go to infinity, the average number of message symbols recovered at $\text{U}_1$ from other users can be approximated by $(M-1)k(1-p_{1,l})$, when $l$ goes to infinity (Fig. \ref{FigRatioOfRec2}).
\subsubsection{Code Performance at the Destination in the General M-user CMAC}
Following a similar approach as in the two-user case, we divide $\text{U}_i$'s message symbols into $L_i$ different parts and let $P^{(i)}_j$ denote the $j^{th}$ part of $\text{U}_i$'s message symbols recovered in $\text{TF}_j$. Then, we can map each part of the message to a separate group of OR nodes. This leads to $N_O=\sum_{i=1}^{M}L_i$ different types of OR nodes, namely Type-$X_1$, Type-$X_2$, ..., and Type-$X_L$ OR nodes. Furthermore, each coded symbol may connect to a different set of OR nodes, which leads to a total of  $N_A=2^L-1$ different types of AND nodes, and each type of AND nodes is represented by a vector of dimension $N_A$. We also consider $N_O$ different AND-OR trees, $T_{l,1}$, $T_{l,2}$,..., and $T_{l,L}$, where the root of $T_{l,j}$ is a Type-$X_j$ OR nodes. Similar to Lemma \ref{PCClemmaFinal},  the probability that a message symbol of each user is not recovered after $l$ iterations at the destination can be calculated by using (\ref{PCCLemma}), (\ref{betaPCC}), and (\ref{deltaPCC}), via substituting $L$ with $N_O=\sum_{i=1}^{M}L_i$.
\subsection{Degree Distribution Optimization of the PCC Scheme }
By using Lemma \ref{PCClemmaFinal}, we can define an optimization problem to find the optimum degree distribution, which minimizes the error probability at the destination. However, the recovery probability of message symbols at the destination depends on channel erasure probabilities. We need to obtain an optimal degree distribution which is independent of the channel erasure probability.

Let us assume that part $P^{(j)}_i$ of $\text{U}_j$ is the message symbols of an equivalent user $\text{U}_{P^{(j)}_i}$. Then, we have $L$ equivalent users $\text{U}_{P^{(j)}_i}$, $i=1,...,L_j$, $j=1,...,M$, where message lengths of $\text{U}_{P^{(j)}_i}$ is $p^{(j)}_i$. Since part $P_i^{(j)}$ of $\text{U}_j$ is recovered at other users in $\text{TF}_i$, $p_i^{(j)}$ can be roughly estimated by $s^{(j)}-s^{(j-1)}$. Also, due to symmetry the $i^{th}$ part of each user's message will be decoded at the same time at the destination. Note that the $ i^{th}$ part of each user's message has the same length due to the symmetry, but, they can have possibly any configuration due to the randomness of the channels. Similar to the optimization problem proposed for the FCC scheme, we can determine the degree distribution of the coded symbols at the destination when some users' message symbols are already known at the destination. More specifically, in $\text{TF}_{R+1}$ each user generates coded symbols from $k+(M-1)s^{(R)}$ message symbols. As a result, when the destination knows message symbols of $\text{U}_{P_i^{(j)}}$, $i=1,2,...,R$ and $j=1,2,...,M$, it removes all edges connected to these symbols and thus the degree distribution of coded symbols will be
\begingroup\makeatletter\def\f@size{9}\check@mathfonts
\def\maketag@@@#1{\hbox{\m@th\small\normalfont#1}}%
\begin{align}
\label{newdegreePCC}
\Delta^{(R)}_d=\sum_{l=0}^{(M-1)s^{(R)}}\Omega_{d+l}\frac{\dbinom{Ms^{(R)}}{l}\dbinom{M(k-s^{(R)})}{d}}{\dbinom{Mk}{l+d}}, \end{align}\endgroup
for $d=0,1,...,k$. Following the similar optimization approach in \cite{CoopLowMACtrans}, and that for the FCC scheme, to ensure that the destination can decode $M(k-s^{(R)})$ remaining message symbols at a minimum overhead, the degree distribution of coded symbols, $\Delta^{(R)}(x)$ should satisfy the following condition
\begingroup\makeatletter\def\f@size{9}\check@mathfonts
\def\maketag@@@#1{\hbox{\m@th\small\normalfont#1}}%
\begin{align}
\label{PCCakhar}
\Delta^{(R)'}(x)\ge-r_R\text{ln}\left(1-x-c\sqrt{\frac{1-x}{M(k-s^{(R)})}}\right),
\end{align}\endgroup
for $x\in[0,1-\delta]$. By using (\ref{newdegreePCC}) and (\ref{PCCakhar}), the linear optimization problem can be summarized as follows.
\begingroup\makeatletter\def\f@size{9}\check@mathfonts
\def\maketag@@@#1{\hbox{\m@th\small\normalfont#1}}%
\begin{align}
\nonumber \text{maximize}~~ \sum_{j=1}^{L}r_j
\end{align}\endgroup
subject to
\begingroup\makeatletter\def\f@size{9}\check@mathfonts
\def\maketag@@@#1{\hbox{\m@th\small\normalfont#1}}%
\begin{eqnarray}
\label{OptProbPCC}
(i). &\displaystyle\sum_{d=1}^k\sum_{l=0}^{Ms^{(j)}}d\Omega_{d+l}\frac{\dbinom{Ms^{(j)}}{l}\dbinom{M(k-s^{(j)})}{d}}{\dbinom{Mk}{l+d}}x^{d-1}\nonumber\\
\nonumber&\ge-r_j\text{ln}\left(1-x-c\sqrt{\frac{1-x}{M(k-s^{(j)})}}\right),~1\le j\le L,\\
\nonumber (ii). &\quad \displaystyle\sum_{d=1}^k\Omega_d=1 ~,~\\
\nonumber (iii).  &\quad \displaystyle0\le\Omega_d\le1, ~~~~~~~d=1,2,...,k,
\end{eqnarray}\endgroup
where in (i), $x\in[0,1-\delta]$. Since $s^{(i)}$ depends on $\Omega(x)$, this cannot be determined without knowing the optimum degree distribution function. In order to find the optimum degree distribution by using a linear programming algorithm, we first take the initial value for each $s^{(i)}$. As in $TF_i$ each user has received $iN$ coded symbols from each of all the other users, the maximum number of recovered information symbols from each user, is roughly estimated to be $iN$. For simplicity this value is considered as the initial value of $s^{(i)}$. Note that the initial values can be any values as long as it is an increasing function of $i$ and satisfies the condition of $s^{(i)}\le Ni$. Based on these values, we find the degree distribution. In the next iteration, we estimate the value of $s^{(i)}$ by using (\ref{OptDegrOpt}) and the degree distribution obtained in the previous iteration, and find the optimum degree distribution based on the new value of $s^{(i)}$. This process continues until the value of $s^{(i)}$ and so $\Omega(x)$ do not change in the consecutive iterations. In practice, after only three iterations, the optimization problem converges to its solution. Table \ref{OptDegPCC} shows some optimized degree distributions obtained by this method for different values of $M$ and $N$, when $k=10000$.

\section{Simulation Results}
\label{simulationresults}
In this section, we compare the performance of the proposed FCC and PCC schemes with the conventional no-cooperation and perfect-cooperation schemes. The upper bounds on the average system throughput of the PCC and FCC schemes are also provided. In the no-cooperation scheme, each user only generates Raptor coded symbols from its own message symbols without cooperation with other users, and transmits them to the destination. In the perfect-cooperation scheme, each user's message is assumed to be perfectly known to the other users, so each user generates coded symbols from all users' messages and then transmits them in its allocated TS in each TF. Therefore, in the perfect-cooperation scheme, the destination receives Raptor coded symbols from all the users which have been generated from message symbols of all users using different Raptor codes.
\subsection{Upper Bounds for the Average System Throughput}
Due to space limit, we only provide the upper bound on the average system throughput for the 2-user CMAC. Similar method can be applied for the general M-user case.
\begin{lemma}
\label{UppFCCPCC}
Let $\eta_{FCC}$ and $\eta_{PCC}$ denote the average system throughput of the FCC and PCC schemes, respectively, then  for the 2-user CMAC, $\eta_{FCC}$ and $\eta_{PCC}$ are upper bounded as follows:
\begingroup\makeatletter\def\f@size{9}\check@mathfonts
\def\maketag@@@#1{\hbox{\m@th\small\normalfont#1}}%
\[\eta_{FCC} \le \left\{
  \begin{array}{l l}
    \frac{2-e_1-e_2}{2} & \quad \text{if}~ 0\le e\le\text{min}\{e_1,e_2\},\\
        1-\text{max}\{e_1, e_2\} & \quad \text{if}~ \text{max}\{e_1, e_2\}\le e\le 1,\\
        \frac{(1-e)(2-e_1-e_2)}{2-e-\text{min}\{e_1,e_2\}} & \quad \text{otherwise},
  \end{array} \right.\]\endgroup
and
\begingroup\makeatletter\def\f@size{9}\check@mathfonts
\def\maketag@@@#1{\hbox{\m@th\small\normalfont#1}}%
\begin{align}
\eta_{PCC}\le\frac{k}{(L+1)N},
\end{align}\endgroup
where $L=\text{max}\{L_1,L_2\}$ and $L_1$ is the smallest integer $M$, which satisfy the following condition:
\begingroup\makeatletter\def\f@size{9}\check@mathfonts
\def\maketag@@@#1{\hbox{\m@th\small\normalfont#1}}%
\begin{align}
\label{PCCshart}
N(1-e_1)\left(1+\sum_{i=2}^{M}\frac{k-s_i}{k}\right)+N(1-e_2)\sum_{i=1}^{M}\frac{s_i}{k}\ge k,
\end{align}\endgroup
where $s_i$ is the number of recovered information symbols at each user in $TF_i$. Also, $L_2$ is the smallest integer $M$, which satisfy (\ref{PCCshart}), when $e_1$ and $e_2$ are replaced by $e_2$ and $e_1$, respectively.
\end{lemma}
The proof of this lemma is provided in Appendix \ref{ProofUpp}.
\subsection{Symmetric Case}
\begin{table}[t]
\caption{Degree distributions for various values of $M$ and $N$;  $\mu$ is the average degree of a coded symbol and $\delta=0.01$}
\label{OptDegPCC}
\centering
\scriptsize
\begin{tabular}{|c||c|c||c|c||c|c|}
\hline
$M$&\multicolumn{2}{|c||}{2}&\multicolumn{2}{|c||}{3}&\multicolumn{2}{|c|}{4}\\
\hline
\hline
$N/k$&0.1&0.05&0.1&0.05&0.1&0.05\\
\hline
\hline
$\Omega_1$& 0.0069&0.0069&0.0057&0.0057&0.0049&0.0049\\
\hline
$\Omega_2$&0.4898&0.4889&0.4907&0.4899&0.4913&0.4905\\
\hline
$\Omega_3$&0.1656&0.1691&0.1660&0.1686&0.1661&0.1680\\
\hline
$\Omega_4$&0.0883&0.0743&0.0883&0.0769&0.0883&0.0799\\
\hline
$\Omega_5$&&0.0224&&0.0182&&0.0135\\
\hline
$\Omega_6$&0.1169&0.1050&0.1172&0.1077&0.1173&0.1106\\
\hline
$\Omega_{13}$&0.0666&0.0693&0.0659&0.0666&0.0653&0.0644\\
\hline
$\Omega_{14}$&0.0207&0.0187&0.0214&0.0210&0.0220&0.0230\\
\hline
$\Omega_{50}$&0.0447& 0.0451&0.0446&0.0448& 0.0445&0.0448\\
\hline
\hline
$\mu$&5.93&5.95&5.92&5.94&5.92&5.94\\
\hline
\end{tabular}
\end{table}
\normalsize
\begin{figure*}
\centering
\mbox{
\subfigure[$e_1=e_2$ and the inter-user channel is error free.]{\includegraphics[width=8.5cm, height=5.4cm]{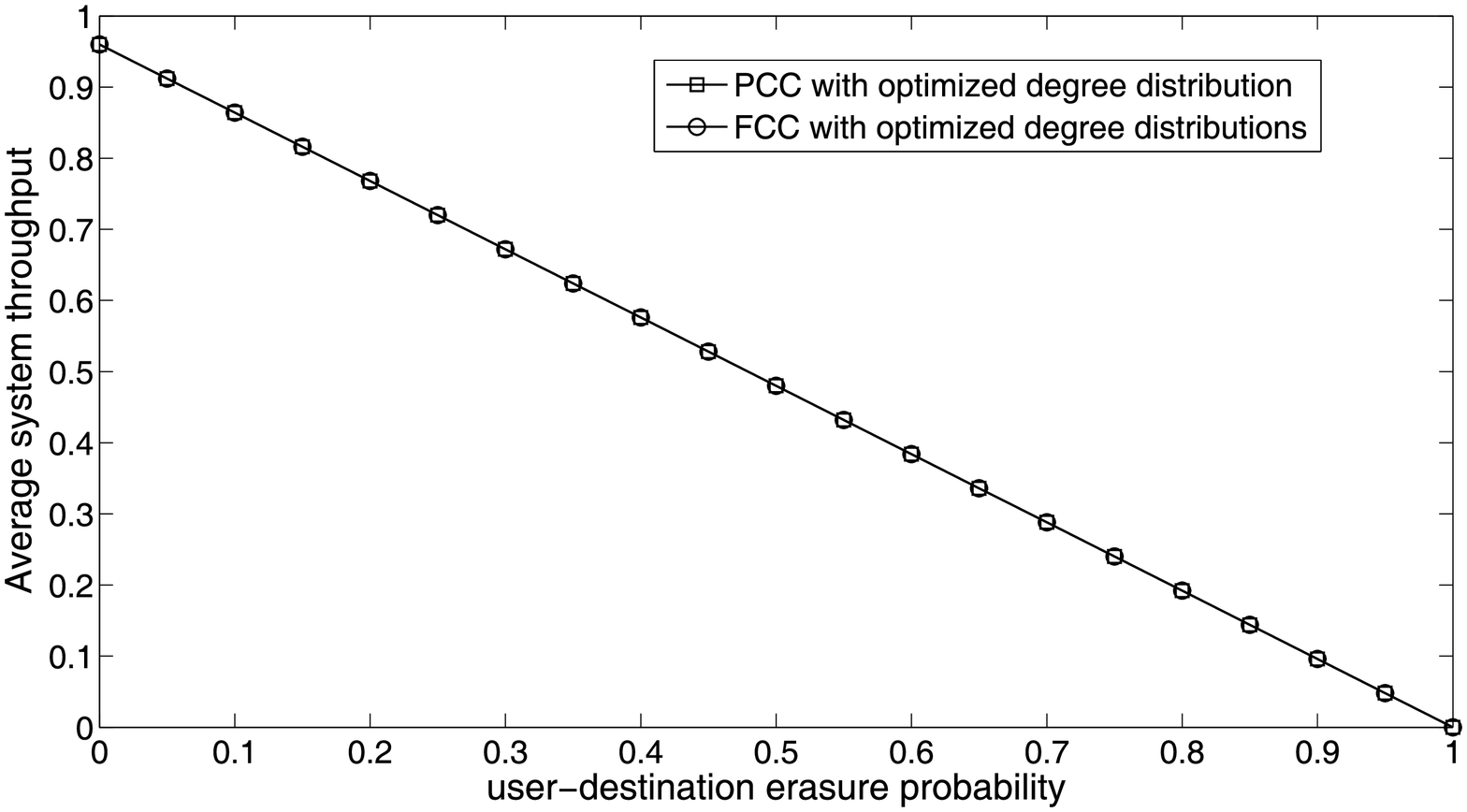}
\label{Symmetric2user}
}
\subfigure[$e_1=0.3$ and $e_2=1$.]{\includegraphics[width=8.5cm, height=5.4cm]{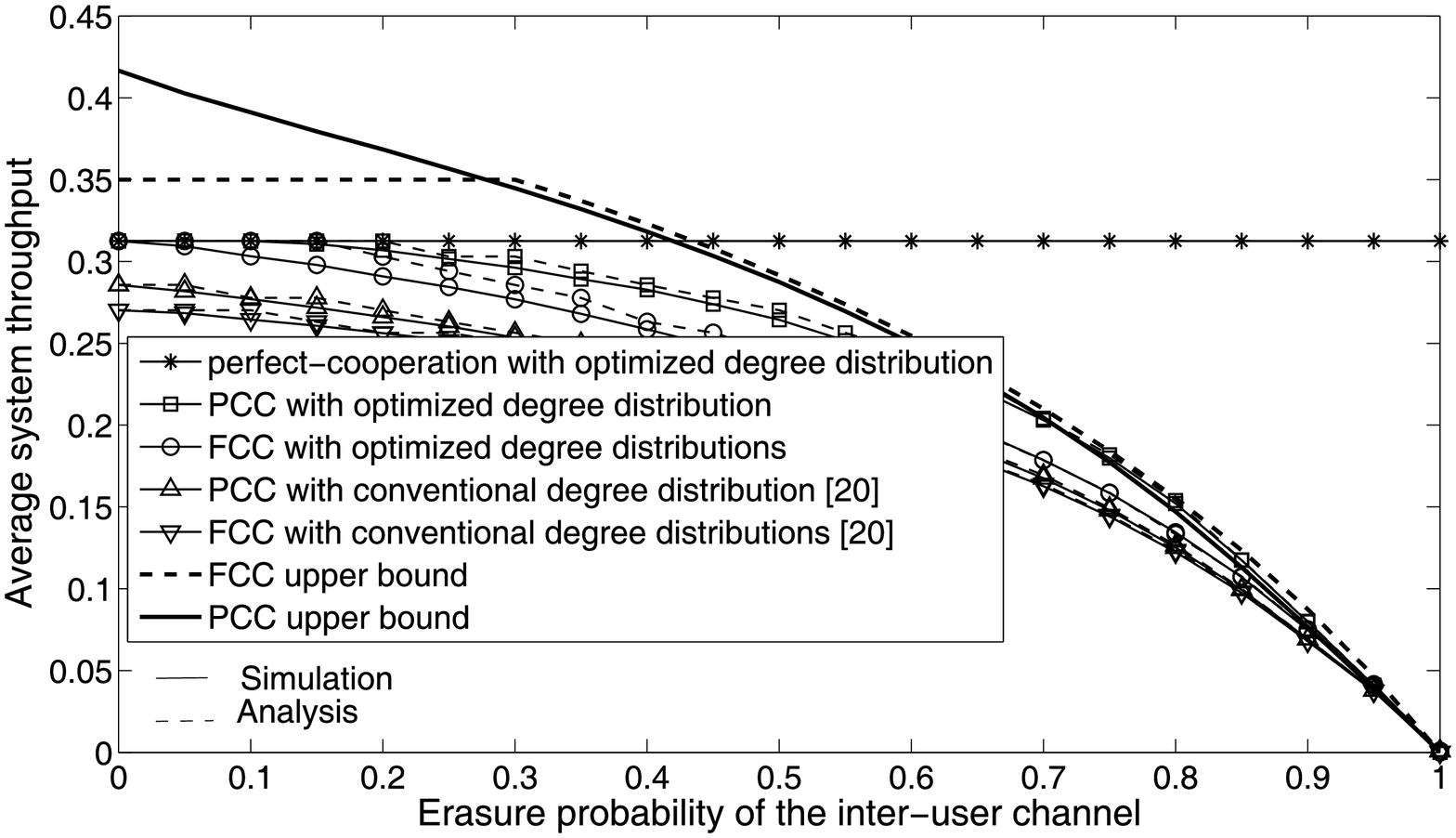}
\label{relaycase}
}
}
\mbox{
\subfigure[$e_1=0.2$ and $e_2=0.8$.]{\includegraphics[width=8.5cm, height=5.4cm]{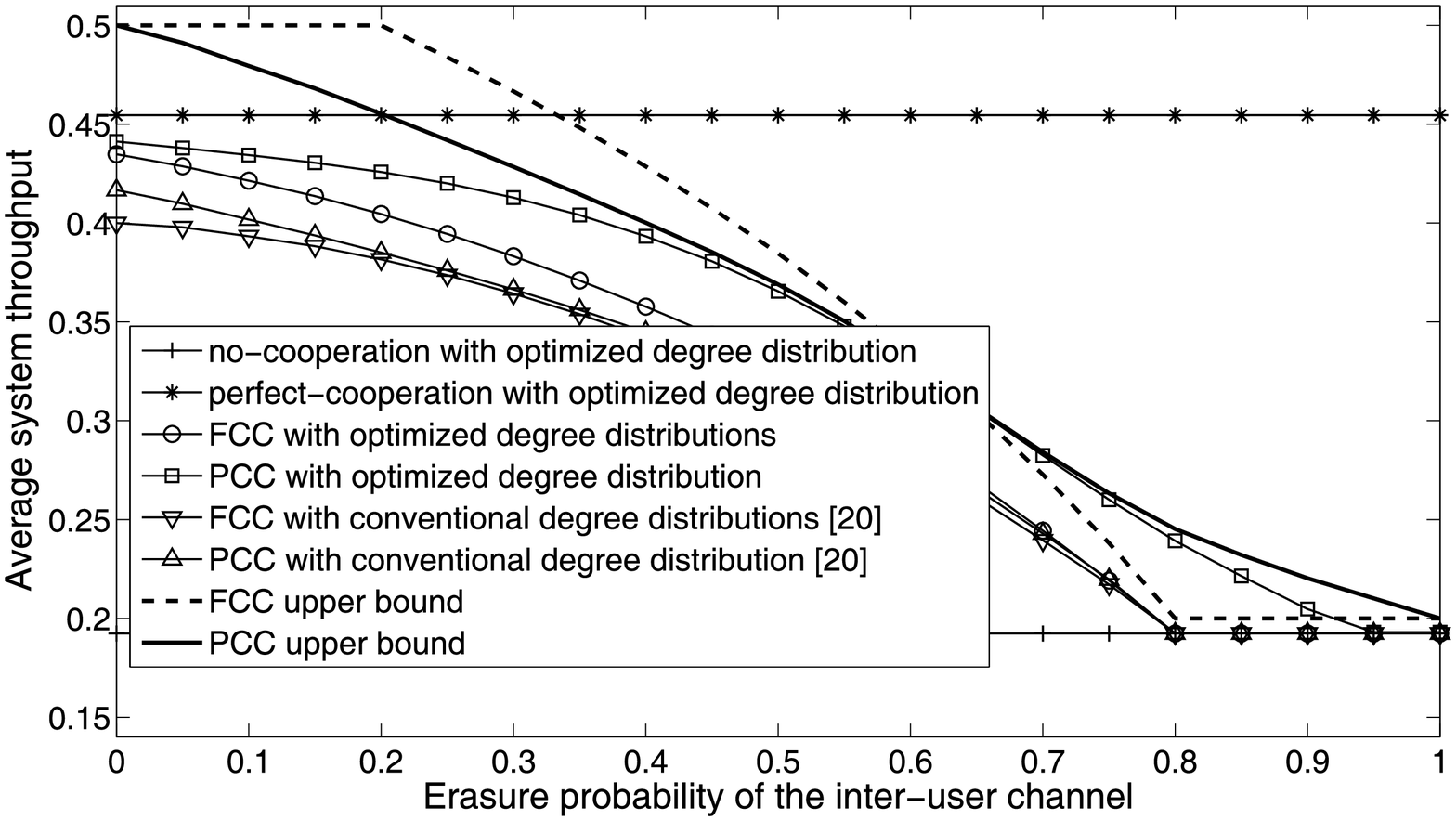}
\label{e102e208}
}
\subfigure[$e_1=0.2$ and $e_2=0.6$.]{\includegraphics[width=8.5cm, height=5.4cm]{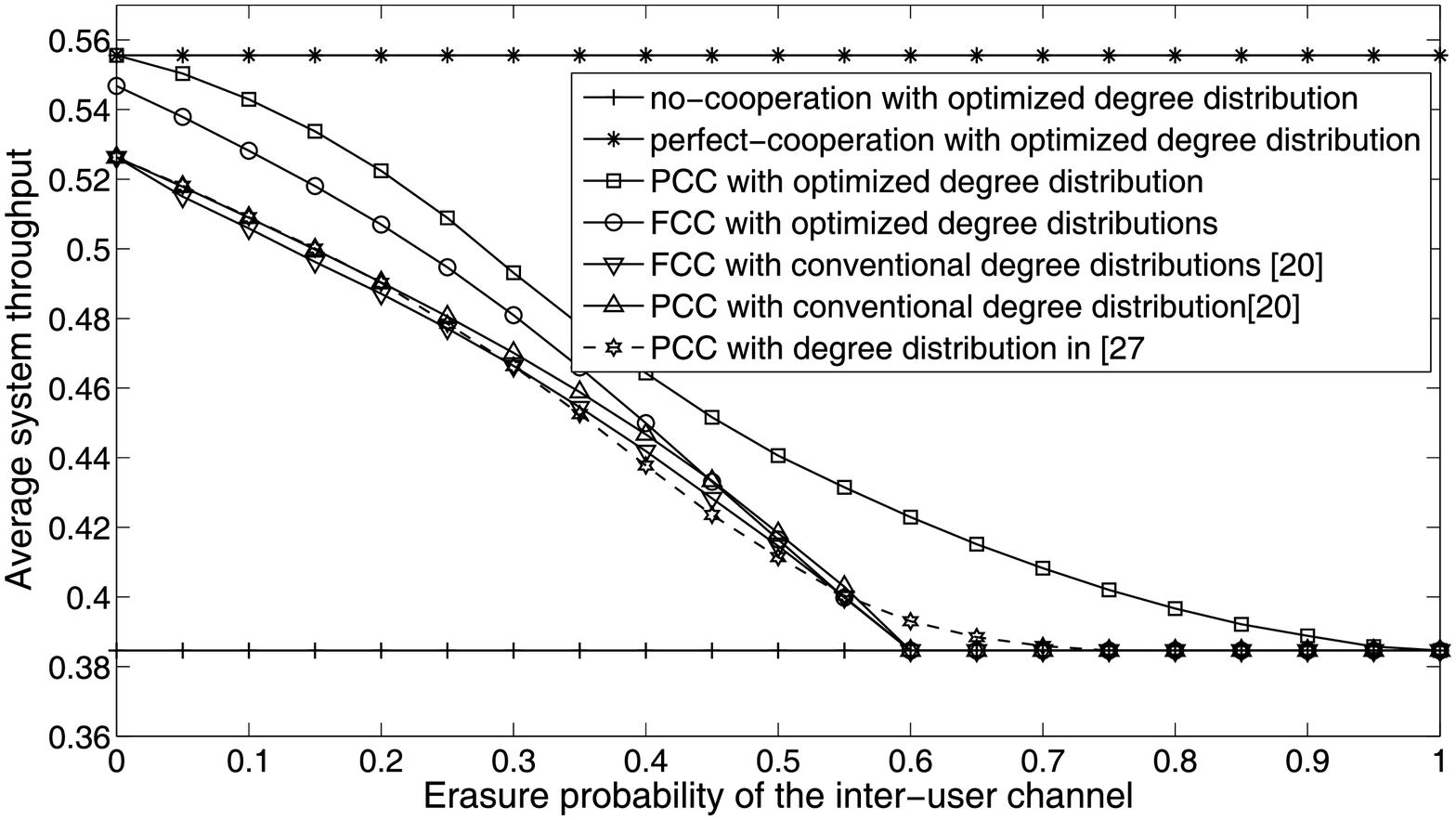}
\label{e102e206}
}
}
\caption{Average system throughput versus the erasure probability for a 2-user CMAC when $k=10000$ and $N=1000$.}
\label{2user}
\end{figure*}
Let us first investigate a 2-user CMAC and assume that the erasure probability of the channel between each user and the destination is the same, i.e., $e_1=e_2$.  In the no-cooperation scheme, the destination decodes $\text{U}_1$'s information symbols as soon as it receives at least $k(1+ \delta_k)$ coded symbols, where $\delta_k$ is the average overhead required to ensure a successful decoding at the destination. Since the symbols transmitted from $\text{U}_1$ may be erased, $\text{U}_1$ should transmit at least $k(1+\delta_k)/(1-e_1)$ coded symbols. Similarly,
$\text{U}_2$ should send $k(1+\delta_k)/(1-e_2)$ coded symbols. If $e_1=e_2=e$, then, the users should send overall $2k(1+\delta_k)/(1-e)$ coded symbols to the destination to guarantee that both users' messages can be decoded successfully.

In the perfect-cooperation scheme, the destination needs to receive at least $2k(1+\delta_{2k})$ coded symbols in order to fully decode both users' messages, where $\delta_{2k}$ is the average overhead required to ensure a successful decoding of $2k$ message symbols at the destination. Thus, users should send overall $2k(1+\delta_{2k})/(1-e)$ coded symbols ($e_1=e_2=e$). Since, $\delta_k$ and $\delta_{2k}$ go to zero when $k$ goes to infinity \cite{Luby}, the total number of symbols required for a successful transmission in both no-cooperation and perfect-cooperation schemes will be the same when $k$ is relatively large. Therefore, in the symmetric case, for a relatively large $k$, cooperation between the users cannot improve error performance at the destination. Fig. \ref{2user}-a shows the average system throughput versus $e_1$ when $e_1=e_2$ for a 2-user CMAC. Here, the throughput is defined as the ratio of the total number of information symbols and that of coded symbols transmitted by all users to ensure that all information symbols are completely decoded at the destination.

In our simulations, all messages are of length $n=9500$ bits and are precoded using a rate $0.95$ LDPC code to obtain $k=10000$ message bits. $N$ is $1000$ and the maximum number of LT decoding iterations at the users and the destination is $100$.
\subsection{Asymmetric Cases}
Fig. \ref{2user}-b, c, and d show the average system throughput versus the erasure probability of the inter-user channel for asymmetric cases in a 2-user CMAC. For clear illustration of results, we show the analytical results only in Fig. \ref{2user}-b, which show an excellent agreement with the simulation results. Results show that the PCC scheme always outperforms the FCC scheme in simulated scenarios. It can be noted that the FCC scheme performs very similar to the no-cooperation scheme when the erasure probability of the inter-user channel is larger than that of users to destination channels. This is because in the FCC scheme, each user needs to wait until it can fully recover the other user's message before starting the cooperation phase. Furthermore, the PCC scheme performs very close to the perfect-cooperation scheme when the inter-user channel has a lower erasure probability.

In the FCC scheme, different users need to select degree distribution $\Phi^{(m+1)}(x)$ when it successfully decodes $m$ users' message symbols. However, in PCC we use a fixed degree distribution during the whole transmission. It can be clearly seen from Tables I and III that the average degree of degree distribution optimized for the PCC scheme is about 6 while that for the FCC scheme is about 8 which further increases with the number of users. This means that the encoding process of the FCC scheme is more complex than that of the PCC scheme.

Fig. \ref{2user}-c and d also show that the codes with the optimized degree distributions considerably outperform those with the conventional degree distribution functions optimized for the point to point channels \cite{Raptor}. More specifically, for the PCC scheme, the derived optimal degree distribution can bring on average $25\%$ improvement in overall throughput, compared to the conventional degree distributions, when $e_1=0.2$ and $e_2=0.8$. This is because the PCC scheme requires a degree distribution with both partial and full decoding performance, but LT codes with the conventional  degree distributions \cite{Raptor} only guarantee the good full decoding performance but cannot provide a good partial decoding performance \cite{ImprovedInterPerfRateless}.

In Fig. \ref{2user}-d, we also compared throughput performance of PCC with optimized degree distributions and that with the conventional distributions obtained in  \cite{RateIntermediate}. As can be seen in this figure, our optimized degree distribution can achieve a considerably higher average system throughput than the one obtained in \cite{RateIntermediate}. This is because the degree distribution in \cite{RateIntermediate} was not designed for the PCC scheme and thus, cannot achieve the optimum performance.
\begin{figure*}[t]
\centering
\mbox{
\subfigure[$e_1=0.2$, $e_2=0.6$]{\includegraphics[width=8.5cm, height=5.4cm]{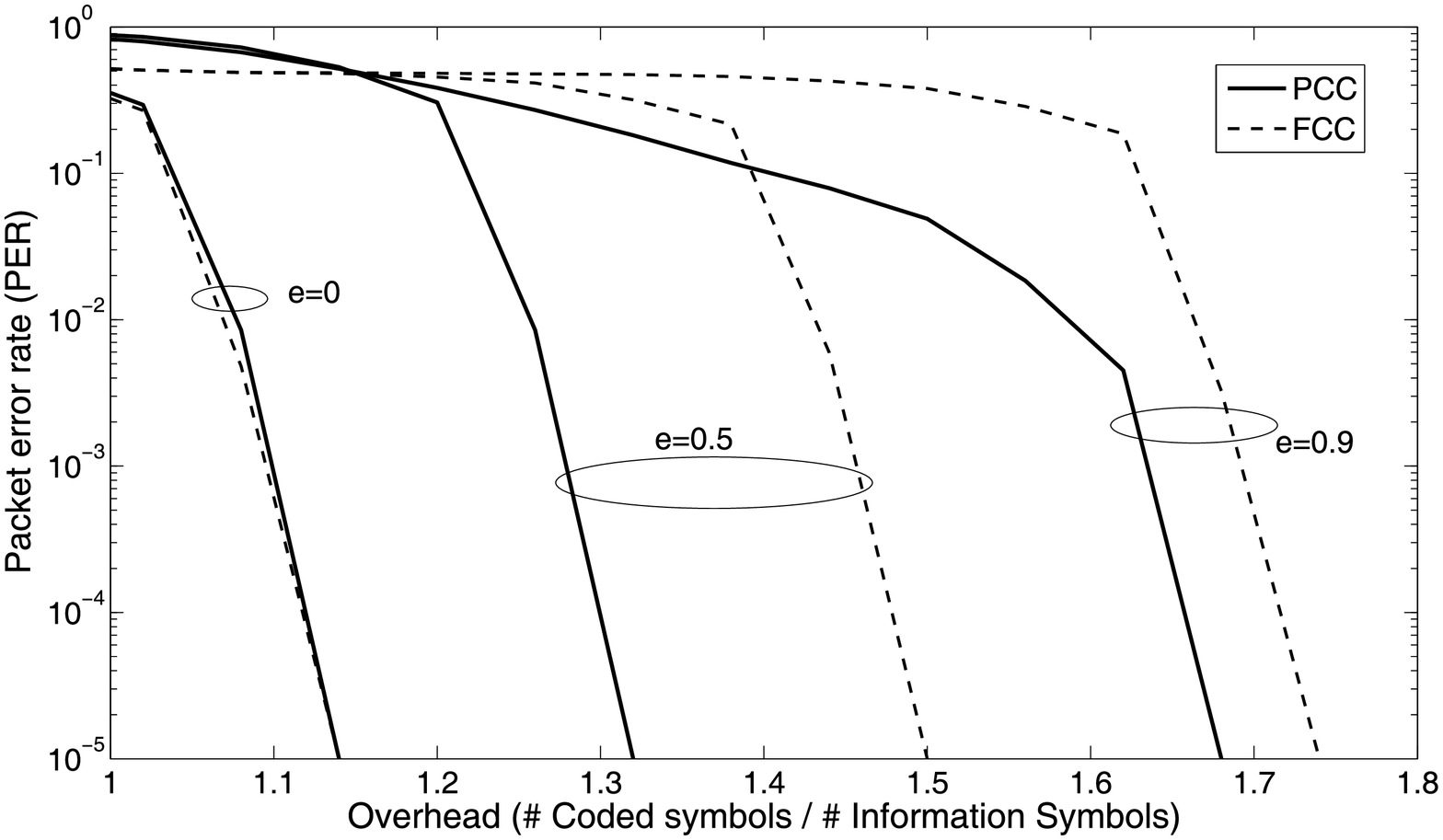}
\label{Beroverhead1}
}
\subfigure[$e_1=0.2$, $e_2=0.8$]{\includegraphics[width=8.5cm, height=5.4cm]{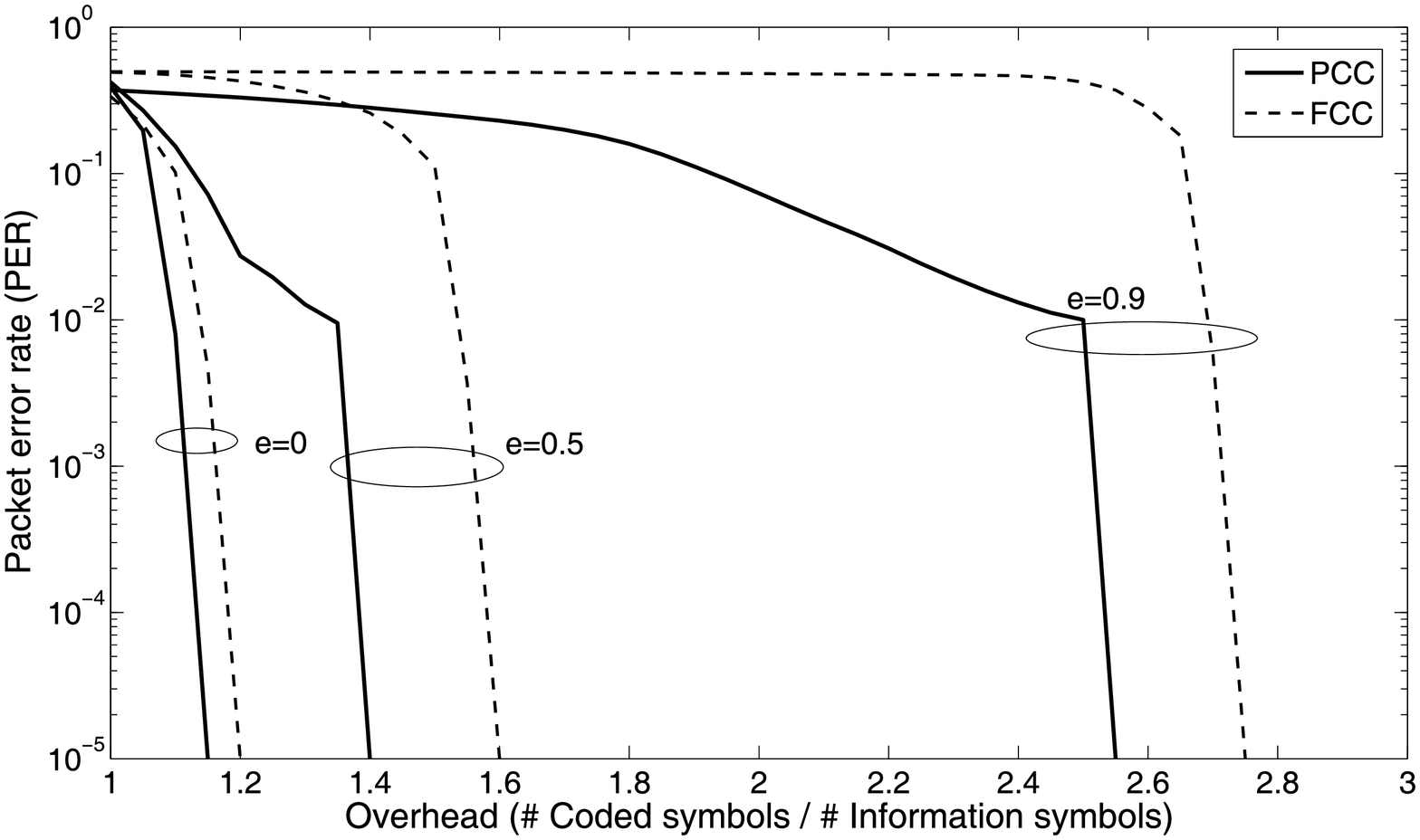}
\label{Beroverhead2}
}
}
\caption{Packet error rate (PER) versus the overhead. Overhead is defined as the ratio of the number of coded symbols and that of information symbols.}
\label{BERoverhead}
\end{figure*}

Fig. \ref{BERoverhead} shows the packet error rate (PER) versus the overhead for a 2-user CMAC for different erasure probability of the inter-user channel. As can be seen in this figure, the PCC scheme outperforms the FCC scheme in terms of PER performance for a certain number of transmission overhead. This is because the destination requires a lower number of coded symbols to completely decodes both users information symbols. Note that overhead is a common performance measure of rateless codes and defined as the ratio of the number of coded symbols and that of information symbols.

\begin{figure*}[t]
\centering
\mbox{
\subfigure[$e_1=0.2$, $e_2=0.4$, $e_3=0.6$, $e_4=0.8$]{\includegraphics[width=8.5cm, height=5.4cm]{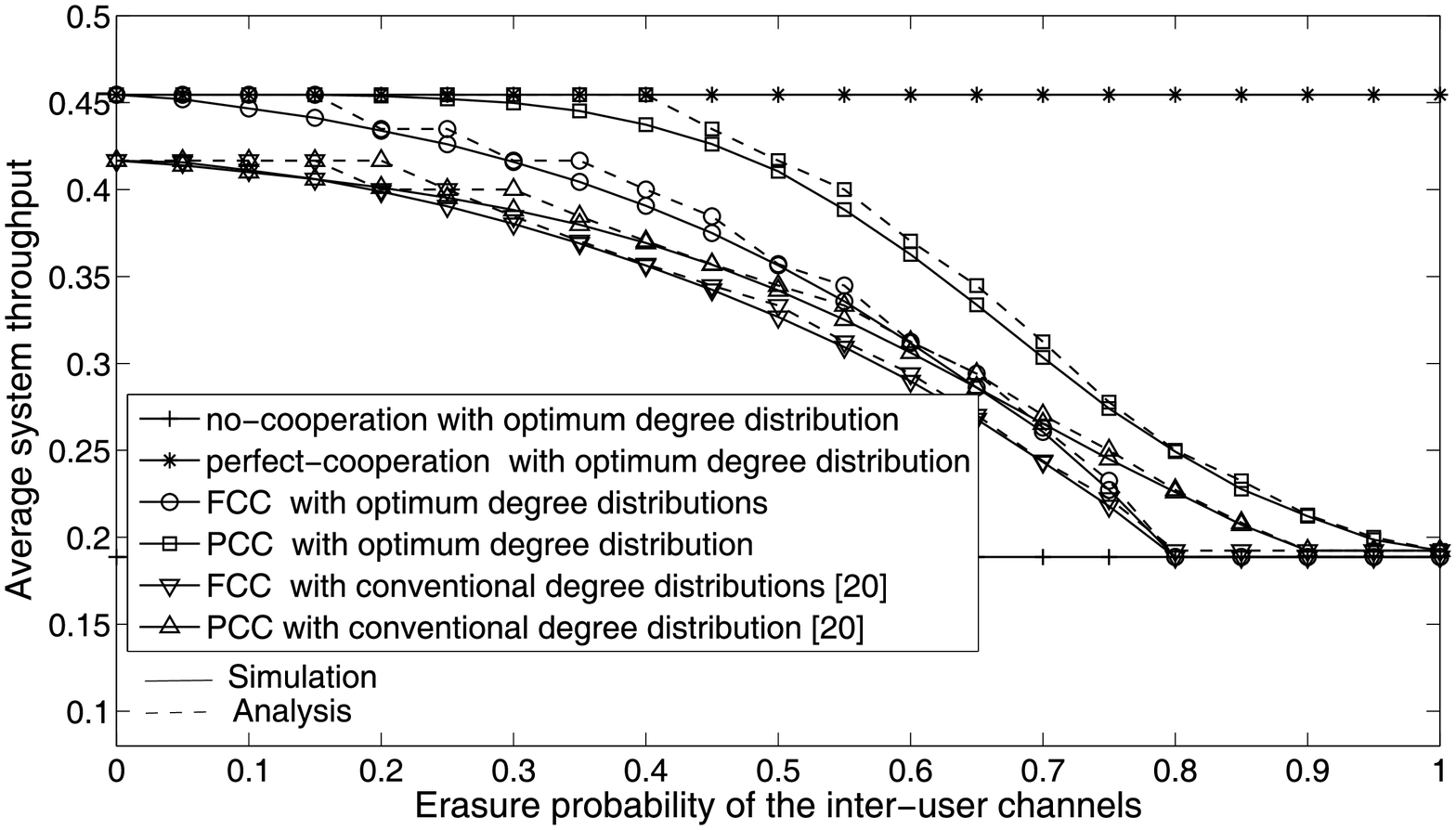}
\label{4user2468}
}
\subfigure[$e_1=0.2$, $e_2=0.4$, $e_3=0.6$, $e_4=0.6$]{\includegraphics[width=8.5cm, height=5.4cm]{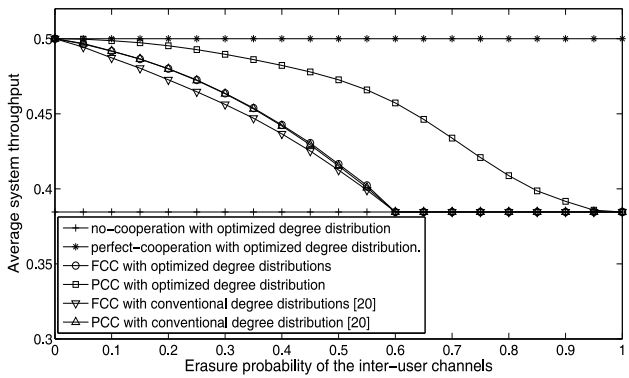}
\label{4user2466}
}
}
\caption{Average system throughput versus the erasure probability of the inter-user channel for a 4-user CMAC when $k=10000$ and $N=1000$. All inter-user channels have the same erasure probability, $e$.}
\label{4user}
\end{figure*}

Fig. \ref{4user} shows the average system throughput versus the erasure probability of the inter-user channel for a 4-user CMAC. We assume that all inter-user channels have the same erasure probability. In Fig \ref{4user}-a we set  $e_1=0.2$, $e_2=0.4$, $e_3=0.6$, and $e_4=0.8$, and in Fig. \ref{4user}-b we have $e_1=0.2$, $e_2=0.4$, $e_3=0.6$, and $e_4=0.6$. It can be noted from the figures that the PCC scheme outperforms the FCC scheme in both cases. For example in Fig. \ref{4user}-b, the PCC scheme can bring about 11\% throughput gain compared to the FCC scheme at the inter-user erasure probability of 0.45 and the gain increases to 20\% when the inter-user erasure probability is 0.6. Also, 25\% (in Fig. \ref{4user}-a) and 20\% (in Fig. \ref{4user}-b) performance gain can be achieved on average for the PCC scheme by using optimized degree distributions compared to the PCC scheme with conventional degree distributions \cite{Raptor}.

\section{Conclusion}
In this paper, we proposed new rateless coded cooperation schemes with partially coded cooperation (PCC) and fully coded cooperation (FCC) for multi-user multiple-access channels. The performance of both schemes have been analyzed using AND-OR tree analysis and validated by simulations. For each scheme, we formulated a linear programming optimization problem to find optimum degree distributions to maximize the average system throughput. Both simulation and analytical results show that the PCC scheme outperforms the FCC scheme in terms of average system throughput. Moreover, the PCC and FCC schemes with optimized degree distributions considerably outperform those with the conventional degree distributions originally designed for point-to-point transmission.

\appendices
\section{Proof of Lemma \ref{PCCandortreelemma}}
\label{PCCandortreelemmaProof}
Each Type-\textbf{V} AND node at depth $2l-1$ calculates its value by performing AND operation on children at depth $2l$. By definition, a Type-\textbf{V} AND node has $i_j$ Type-$X_j$ OR children with the probability $\beta_{\textbf{V},\textbf{I}}$ and the value of it will be 1 if its all children are 1. This means that the probability that a Type-\textbf{V} AND node at depth $2l-1$ which has $d$ children has a value of 1 is $\beta_{\textbf{V},\textbf{I}}\prod_{w=1}^{N_O}(1-p_{0,w})^{v_wi_w}$. Summation over all possible $d$ and vectors $\textbf{I}$ gives the probability that a Type-\textbf{V} AND node at depth $2l-1$ has the value of 1. Furthermore, a Type-$X_j$ OR node at depth $2l-2$ have a value of zero if its all children have the value of 0. Since a Type-$X_j$ OR node has $i$ type \textbf{V} AND children with the probability $\delta_{j,\textbf{V},i}$, and by summation over all possible vectors $V$, $p_{1,j}$ will be $\prod_{\textbf{V}\in \mathcal{A}_j^{(N_O)}}\delta_{j,\textbf{V}}\left(1-\sum_{d}\sum_{\textbf{I}\in \mathcal{I}_{\textbf{V},j,d}}\beta_{\textbf{V},\textbf{I}}\prod_{w=1}^{N_O}(1-p_{0,w})^{v_wi_w}\right)$. By repeating this procedure for AND and OR nodes of the trees at different depth up to depth zero, (\ref{PCCandOrLemma}) will be straightforward.

\section{Proof of Lemma \ref{FCCMuserAndOrtree}}
\label{FCCMuserAndOrtreeProof}
Let us assume that $\textbf{Y}\in\mathcal{A}^*(\textbf{V})$ and $q_{Y,d}$ is the probability that a degree-$d$ coded symbol which is generated from message symbols of users in $\textbf{Y}$ is actually a degree-$d$ Type-$\textbf{V}$ AND node. Then, $q_{Y,d}$ is equivalent to the probability of selecting $d$ OR nodes among Type-$X_j$'s OR nodes where $v_j\ne0$. Moreover, a Type-$\textbf{V}$ AND node has at least one Type-$X_j$ OR child, where $v_j\ne0$. Thus $\text{n}(\textbf{V})$ out of $d$ message symbols have to be selected from Type-$X_j$ OR nodes with $v_j\ne0$. This will happen with the probability of $(\frac{1}{\text{n}(\textbf{Y})})^{\text{n}(\textbf{V})}$. Then $d-\text{n}(\textbf{V})$ remaining message symbols are uniformly selected from $\text{n}(\textbf{V})$ out of $\text{n}(\textbf{Y})$ users. This happens with the probability of $(\frac{\text{n}(\textbf{V})}{\text{n}(\textbf{Y})})^{d-\text{n}(\textbf{V})}$. Hence, $q_{Y,d}$ is $(\frac{\text{n}(\textbf{V})}{\text{n}(\textbf{Y})})^{d-\text{n}(\textbf{V})}(\frac{1}{\text{n}(\textbf{Y})})^{\text{n}(\textbf{V})}$. Since a coded symbol is of degree $d$ with probability $\Phi^{(\text{n}(\textbf{Y}))}_d$, and $d$ varies from $\text{n}(\textbf{V})$ to $D$, then the probability that a coded symbol which has been generated from message symbols of users in $\textbf{Y}$ is a Type-$\textbf{V}$ AND node, $q_Y$, will be $\Phi^{*(\text{n}(\textbf{Y}))}(\frac{\text{n}(\textbf{V})}{n(\textbf{Y})})(\frac{1}{\text{n}(\textbf{Y})})^{\text{n}(\textbf{V})}$. In addition, the total number of Type-$\textbf{V}$ AND nodes, $T_\textbf{V}$, can be calculated by summing over all $\textbf{Y}$ in set $A^*(\textbf{V})$. This proves (\ref{NumberofV}).

$\beta_{\textbf{V},\textbf{I}}$ is the probability that a randomly chosen edge in the bipartite graph is connected to a degree $d+1$ Type-$\textbf{V}$ AND node that has $v_ji_j$ Type-$X_j$ OR children and $d=v_1i_1+v_2i_2+...+v_Mi_M$. Since a Type-$\textbf{V}$ AND node is generated from message symbols of users in $\textbf{Y}$ using $\Phi^{(\text{n}(\textbf{Y}))}(x)$ at a probability of $q_YN_\textbf{Y}/T_\textbf{V}$, and the total number of edges connected to these nodes is $q_YN_\textbf{Y}\mu^{(\text{n}(\textbf{Y}))}$, then $q_YN_\textbf{Y}(d+1)\Phi^{(\text{n}(\textbf{Y}))}_{d+1}$ edges are connected to Type-$\textbf{V}$ AND nodes with degree $d+1$. Since message symbols are selected uniformly at random from users in $\textbf{V}$, then (\ref{betaFCCV}) is straightforward.

$\delta_{\textbf{V},i}$ is the probability that a Type-$X_j$ OR node is connected to $i$ Type-$\textbf{V}$ AND nodes. The total number of edges connected to Type-$X_j$ OR nodes is $N_{jT}=\sum_{\textbf{V}\in \mathcal{A}^{*}(\textbf{V})}q_YN_\textbf{Y}\mu^{(\text{n}(\textbf{Y}))}$, which arises from the fact that a coded symbol which has been generated from users in set $\textbf{Y}$ is a Type-$\textbf{V}$ AND node with the probability of $q_Y$. The number of edges connected to a specific Type-$X_j$ OR node is then binomially distributed with parameter $1/ \text{n}(\textbf{V})k$, which can be approximated by a poisson distribution as $e^{-\alpha_\textbf{V}}(\alpha_\textbf{V})^i/i!$, where $\alpha_\textbf{V}=N_{jT}/\text{n}(\textbf{V})k$ and $\delta_{\textbf{V}}(x)=\sum_{i}e^{-\alpha_\textbf{V}}\frac{\alpha_\textbf{V}^i}{i!}x^i=e^{\alpha_\textbf{V}(x-1)}$. This completes the proof.

\section{Proof of Lemma \ref{pldecreaseLI}}
\label{pldecreaseLIProof}
$(1)$. We have $p_1^{(i+1)}=e^{-N^{(i+1)}\Delta_1^{(i+1)}/k}<1=p_0^{(i+1)}$. Suppose that $p_l^{(i+1)}<p_{l-1}^{(i+1)}$.
We need to show that $p_{l+1}^{(i+1)}<p_l^{(i+1)}$, which can  be shown easily using the fact that $\delta^{(i+1)}(.)$ and $\beta^{(i+1)}(.)$ are both increasing functions of their argument.

$(2)$. We have ${\scriptstyle p_0^{(i)}=p_0^{(i+1)}=1}$. We suppose that ${\scriptstyle p_{l-1}^{(i)}\le p_{l-1}^{(i+1)}}$, then we need to show that $p_l^{(i)}\le p_l^{(i+1)}$. First
\begingroup\makeatletter\def\f@size{9}\check@mathfonts
\def\maketag@@@#1{\hbox{\m@th\small\normalfont#1}}%
\begin{align}
\label{firstline}
&\alpha^{(i+1)}\delta^{(i+1)}(x)=\frac{\Delta^{(i+1)'}(x)}{\Delta^{(i+1)'}(1)}\frac{N^{(i+1)}\Delta^{(i+1)'}(1)}{k}\\
&\nonumber\overset{(a)}{=}\sum_{w=1}^{i+1}\frac{N(1-e)}{i+1}\sum_{j=1}^{i+1}\sum_{d=1}^{k}d\Omega_d^{(j)}x^{d-1}\\
&\nonumber=N(1-e)\sum_{j=1}^{i+1}\sum_{d=1}^{k}d\Omega_d^{(j)}x^{d-1}\\
&\nonumber\ge N(1-e)\sum_{j=1}^{i}\sum_{d=1}^{k}d\Omega_d^{(j)}x^{d-1}=\alpha^{(i)}\delta^{(i)}(x),
\end{align}\endgroup
where step (a) follows from substituting (\ref{PCCOutDegFinal}) in (\ref{firstline}) by considering that $ N^{(i+1)}_{2}=(i+1)N(1-e)$. Thus $\alpha^{(i)}\delta^{(i)}(x)$ is an increasing function of $i$. Since we suppose that $p_{l-1}^{(i)}\le p_{l-1}^{(i+1)}$, $(1-p_{l-1}^{(i)})\ge (1-p_{l-1}^{(i+1)})$ and accordingly $\alpha^{(i+1)}\delta^{(i+1)}(1-p_{l-1}^{(i+1)})\ge \alpha^{(i)}\delta^{(i)}(1-p_{l-1}^{(i)})$. This results in $p_l^{(i+1)}\le p_l^{(i)}$.

\section{Proof of Lemma \ref{UppFCCPCC}}
\label{ProofUpp}
Without loss of generality, we assume that $e_1<e_2$. As discussed before, when $e>e_2>e_1$, the destination will be able to recover each user's message before the partner does, so there will be no cooperative process in the FCC scheme.  Accordingly, the total number of TFs required to ensure that both users' messages are decoded at the destination will be at least $N_{TF}=\frac{k}{(1-e_2)N}$. Therefore, the average system throughput will be at most $\frac{2k}{N_{TF}\times 2N}$, which leads to upper bound $\eta_{FCC}\le1-e_2$.

When $e_1<e<e_2$, $\text{U}_1$'s message is decoded at the destination before $U_2$. Let $N_{2}$ denote the number of time frames required to ensure that $\text{U}_2$'s message is decoded at $\text{U}_1$. We have $N_2\ge\frac{k}{N(1-e)}$, and when $\text{U}_2$ transmits $N_2$ messages, the destination has already decoded $\text{U}_1$'s message and also received $N_2\times (1-e_2)$ coded symbols from $\text{U}_2$. The destination then requires at least $k- N_2\times N(1-e_2)$ more coded symbols to fully recover $\text{U}_2$'s message which are sent by both users in at least $\frac{k- N_2\times N(1-e_2)}{N(2-e_1-e_2)}$ time frames. Thus, we have $\eta_{FCC}\le\frac{(1-e)(2-e_1-e_2)}{2-e_1-e}$.

Finally, when $e<e_1<e_2$, both users decode the other user's message before the destination. This means that the destination requires at least $2k$ coded symbols to recover both users' information symbols, which have been sent through at least $\frac{2k}{N(2-e_1-e_2)}$ time frames. Since in each TF the total number of $2N$ coded symbols are transmitted from both users, then we have $\eta_{FCC}\le\frac{2-e_1-e_2}{2}$.

For the PCC scheme, we consider that in $TF_i$, $s_i$ information symbols have been decoded at each user from the other user. Simply, the destination receives $N(1-e_2)$ coded symbols from $\text{U}_2$ in $TF_i$ and $ \frac{k-s_i}{k}$ fraction of them carries information of $\text{U}_2$'s message. Similarly, $\frac{s_i}{k}$ fraction of the coded symbols transmitted from $\text{U}_2$ carries information of $\text{U}_1$. In time frame $L_2$, the total amount of information received from $\text{U}_2$ will be $N(1-e_2)\left(1+\sum_{i=2}^{M}\frac{k-s_i}{k}\right)+N(1-e_1)\sum_{i=1}^{M}\frac{s_i}{k}$.  The destination requires at least $k$ coded symbols carrying information of $U_2$, in order to completely decode it; thus, $L_2$ will be the minimum possible value of $M$ that satisfy (\ref{PCCshart}). Similar calculation can be carried out for $\text{U}_1$.  Therefore, $\eta_{PCC}\le\frac{2k}{(L_2)\times 2N}$.

\bibliographystyle{IEEEtran}
\footnotesize
\bibliography{IEEEabrv,References}

%\enlargethispage{-5in}
\newpage
\begin{IEEEbiography}[{\includegraphics[width=1in,height=1.25in,clip,keepaspectratio]{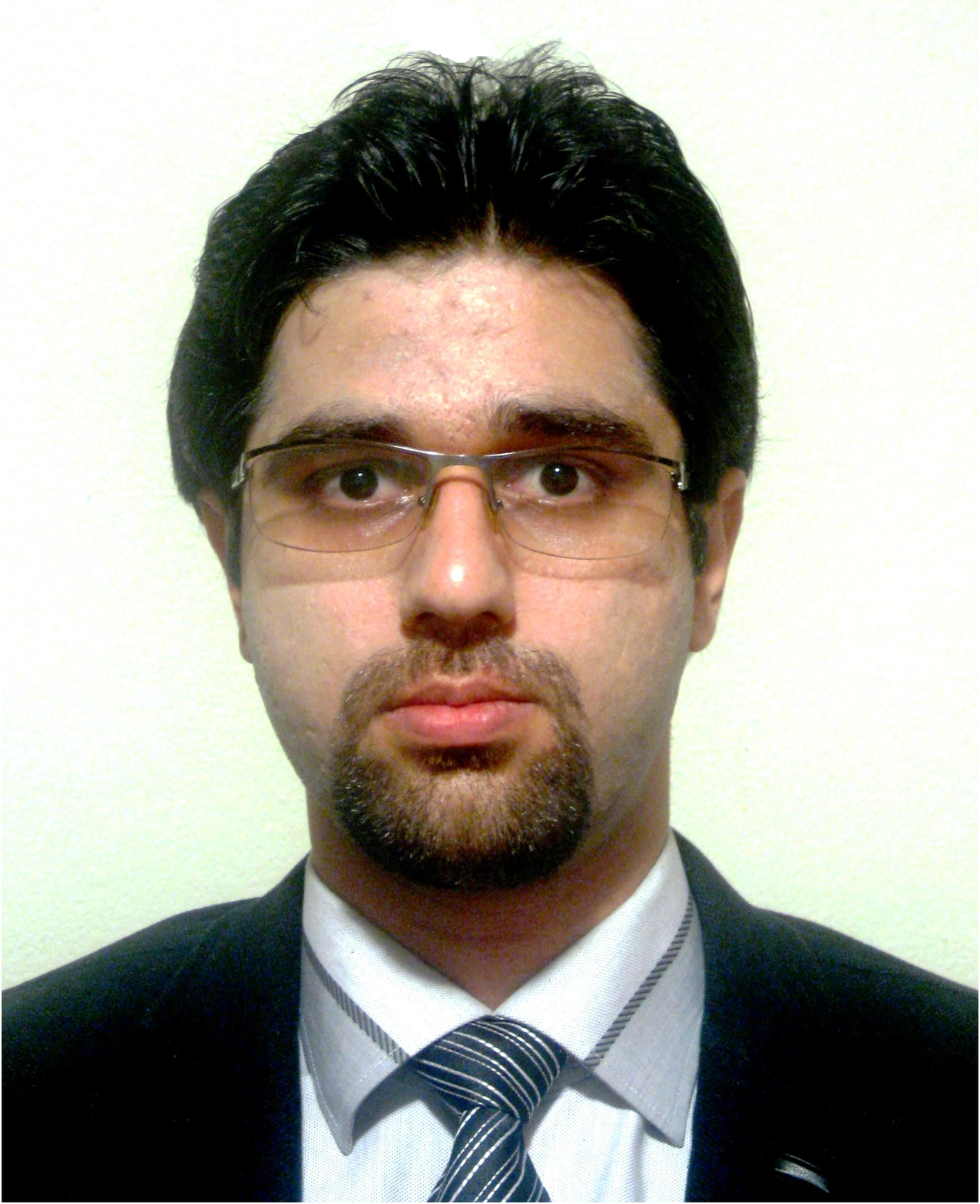}}]{Mahyar Shirvanimoghaddam}
received his B. Sc.
degree with 1'st Class Honours from University of Tehran, Iran, in 2008, and  his M. Sc. Degree with 1'st Class Honours from Sharif University of Technology, Iran, in 2010, both in Electrical Engineering. He is currently
working towards a Ph.D. degree  in Electrical Engineering at The University
of Sydney, Australia.
His research interests include channel coding techniques,
cooperative communications, compressive sensing,
, machine-to-machine communications, and wireless sensor networks. He is a recipient
of University of Sydney International Scholarship
(USydIS) and University of Sydney Postgraduate Award.
\end{IEEEbiography}
\begin{IEEEbiography}[{\includegraphics[width=1in,height=1.25in,clip,keepaspectratio]{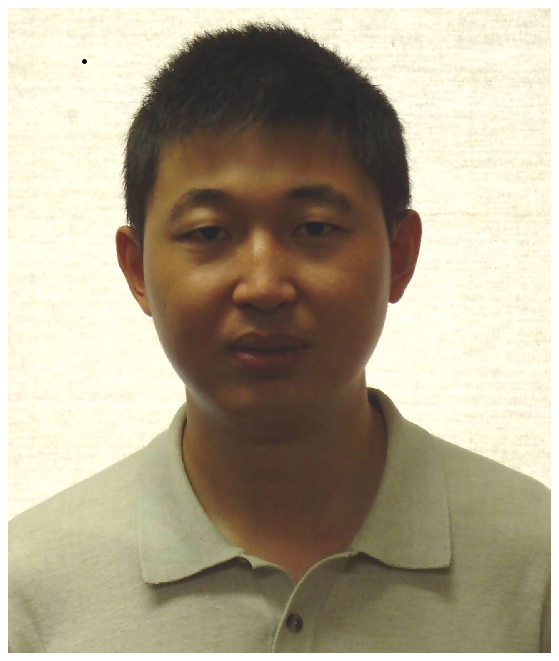}}]{Yonghui Li}
(M'04-SM'09) received his PhD degree in November 2002 from Beijing University of Aeronautics and Astronautics. From 1999 - 2003, he was affiliated with Linkair Communication Inc, where he held a position of project manager with responsibility for the design of physical layer solutions for the LAS-CDMA system. Since 2003, he has been with the Centre of Excellence in Telecommunications, the University of Sydney, Australia. He is now an Associate Professor in School of Electrical and Information Engineering, University of Sydney. He was the Australian Queen Elizabeth II Fellow and is currently the Australian Future Fellow.

His current research interests are in the area of wireless communications, with a particular focus on MIMO, cooperative communications, coding techniques and wireless sensor networks. He holds a number of patents granted and pending in these fields. He is an executive editor for European Transactions on Telecommunications (ETT). He has also been involved in the technical committee of several international conferences, such as ICC, Globecom, etc.
\end{IEEEbiography}
\enlargethispage{-4.9in}
\newpage
\begin{IEEEbiography}[{\includegraphics[width=1in,height=1.25in,clip,keepaspectratio]{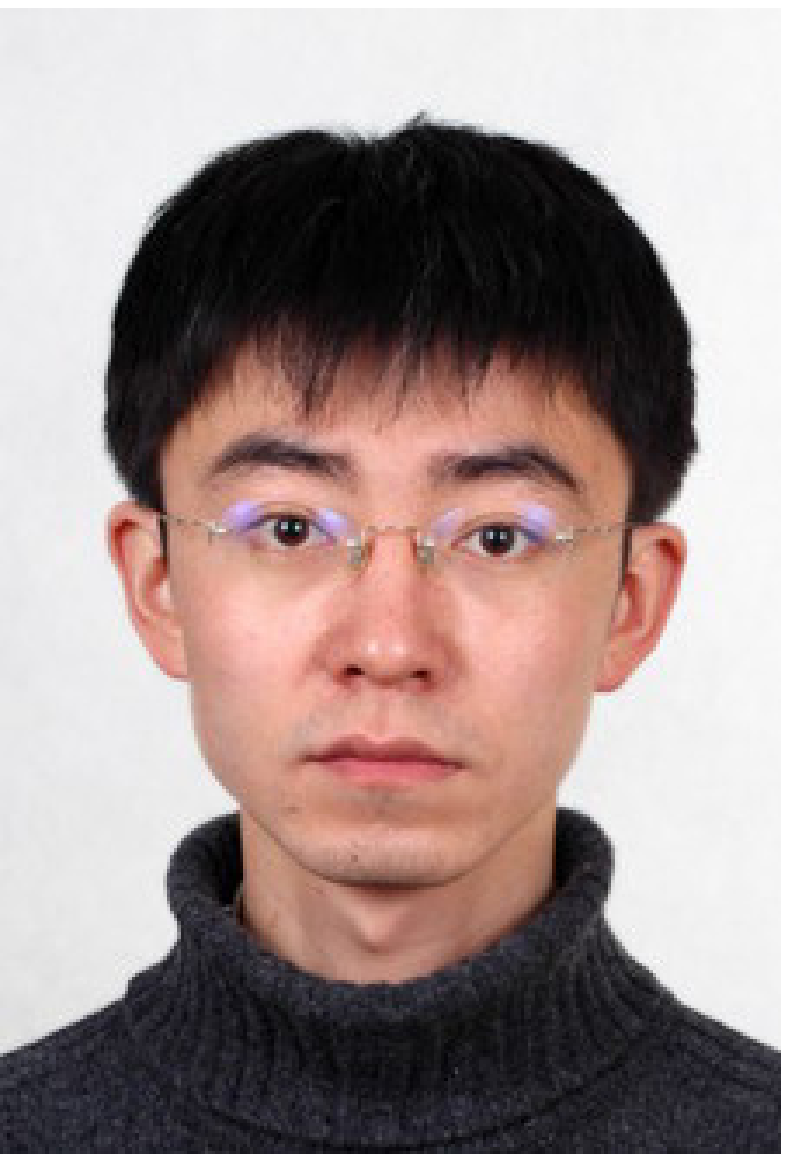}}]{Shuang Tian}
(M'13) received the B. Sc. degree (with highest honors) from Harbin Institute of Technology, China, in 2005, the M. Eng. Sc. (Research) degree from Monash University, Australia, in 2008, and is currently working toward the Ph.D. degree at The University of Sydney, Australia, all in electrical engineering. From 2008 to 2009, he worked as a system and standard engineer with Huawei Technologies Co. Ltd., China. His research interests include channel coding techniques, cooperative communications, 3GPP/WiMAX network protocols, OFDM communications, and optimization of resource allocation and interference management in heterogeneous networks. He is a recipient of University of Sydney Postgraduate Awards.
\end{IEEEbiography}
\begin{IEEEbiography}[{\includegraphics[width=1in,height=1.25in,clip,keepaspectratio]{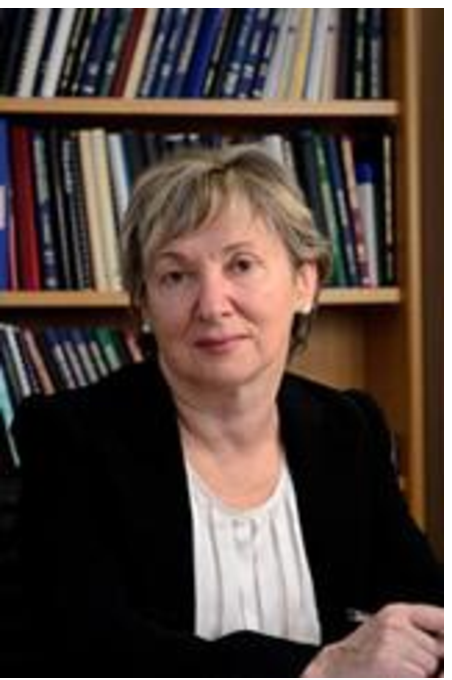}}]{Branka Vucetic}
 (M'83-SM'00-F'03) received the
B.S.E.E., M.S.E.E., and Ph.D. degrees in 1972,
1978, and 1982, respectively, in electrical engineering,
from The University of Belgrade, Belgrade,
Yugoslavia. During her career, she has held various
research and academic positions in Yugoslavia,
Australia, and the UK. Since 1986, she has been
with the Sydney University School of Electrical and
Information Engineering in Sydney, Australia. She
is currently the Director of the Centre of Excellence
in Telecommunications at Sydney University. Her
research interests include wireless communications, digital communication
theory, coding, and multi-user detection.

In the past decade, she has been working on a number of industry sponsored
projects in wireless communications and mobile Internet. She has taught a
wide range of undergraduate, postgraduate, and continuing education courses
worldwide. Prof. Vucetic has co-authored four books and more than two
hundred papers in telecommunications journals and conference proceedings.
\end{IEEEbiography}
\enlargethispage{-5.1in}
\end{document}